\documentclass[11pt]{article}
\usepackage{amsmath,amsthm,latexsym,amssymb,amsfonts,epsfig}

\makeatletter
\@addtoreset{equation}{section}
\makeatother

\def\be{\begin{equation}}
\def\ee{\end{equation}}
\def\ba{\begin{eqnarray}}
\def\ea{\end{eqnarray}}

\newcommand\nn{\nonumber}
\newcommand\q{\quad}
\def\Nl{{\mathchoice
{\setbox0=\hbox{$\displaystyle\rm N$}\hbox{\hbox to0pt
{\kern0.4\wd0\vrule height0.9\ht0\hss}\box0}}
{\setbox0=\hbox{$\textstyle\rm N$}\hbox{\hbox to0pt
{\kern0.4\wd0\vrule height0.9\ht0\hss}\box0}}
{\setbox0=\hbox{$\scriptstyle\rm N$}\hbox{\hbox to0pt
{\kern0.4\wd0\vrule height0.9\ht0\hss}\box0}}
{\setbox0=\hbox{$\scriptscriptstyle\rm N$}\hbox{\hbox to0pt
{\kern0.4\wd0\vrule height0.9\ht0\hss}\box0}}}}
\def\Zl{{\mathchoice
{\setbox0=\hbox{$\displaystyle\rm Z$}\hbox{\hbox to0pt
{\kern0.4\wd0\vrule height0.9\ht0\hss}\box0}}
{\setbox0=\hbox{$\textstyle\rm Z$}\hbox{\hbox to0pt
{\kern0.4\wd0\vrule height0.9\ht0\hss}\box0}}
{\setbox0=\hbox{$\scriptstyle\rm Z$}\hbox{\hbox to0pt
{\kern0.4\wd0\vrule height0.9\ht0\hss}\box0}}
{\setbox0=\hbox{$\scriptscriptstyle\rm Z$}\hbox{\hbox to0pt
{\kern0.4\wd0\vrule height0.9\ht0\hss}\box0}}}}
\def\Ql{{\mathchoice
{\setbox0=\hbox{$\displaystyle\rm Q$}\hbox{\hbox to0pt
{\kern0.4\wd0\vrule height0.9\ht0\hss}\box0}}
{\setbox0=\hbox{$\textstyle\rm Q$}\hbox{\hbox to0pt
{\kern0.4\wd0\vrule height0.9\ht0\hss}\box0}}
{\setbox0=\hbox{$\scriptstyle\rm Q$}\hbox{\hbox to0pt
{\kern0.4\wd0\vrule height0.9\ht0\hss}\box0}}
{\setbox0=\hbox{$\scriptscriptstyle\rm Q$}\hbox{\hbox to0pt
{\kern0.4\wd0\vrule height0.9\ht0\hss}\box0}}}}
\def\Rl{{\mathchoice
{\setbox0=\hbox{$\displaystyle\rm R$}\hbox{\hbox to0pt
{\kern0.4\wd0\vrule height0.9\ht0\hss}\box0}}
{\setbox0=\hbox{$\textstyle\rm R$}\hbox{\hbox to0pt
{\kern0.4\wd0\vrule height0.9\ht0\hss}\box0}}
{\setbox0=\hbox{$\scriptstyle\rm R$}\hbox{\hbox to0pt
{\kern0.4\wd0\vrule height0.9\ht0\hss}\box0}}
{\setbox0=\hbox{$\scriptscriptstyle\rm R$}\hbox{\hbox to0pt
{\kern0.4\wd0\vrule height0.9\ht0\hss}\box0}}}}
\def\Cl{{\mathchoice
{\setbox0=\hbox{$\displaystyle\rm C$}\hbox{\hbox to0pt
{\kern0.4\wd0\vrule height0.9\ht0\hss}\box0}}
{\setbox0=\hbox{$\textstyle\rm C$}\hbox{\hbox to0pt
{\kern0.4\wd0\vrule height0.9\ht0\hss}\box0}}
{\setbox0=\hbox{$\scriptstyle\rm C$}\hbox{\hbox to0pt
{\kern0.4\wd0\vrule height0.9\ht0\hss}\box0}}
{\setbox0=\hbox{$\scriptscriptstyle\rm C$}\hbox{\hbox to0pt
{\kern0.4\wd0\vrule height0.9\ht0\hss}\box0}}}}
\def\Hl{{\mathchoice
{\setbox0=\hbox{$\displaystyle\rm H$}\hbox{\hbox to0pt
{\kern0.4\wd0\vrule height0.9\ht0\hss}\box0}}
{\setbox0=\hbox{$\textstyle\rm H$}\hbox{\hbox to0pt
{\kern0.4\wd0\vrule height0.9\ht0\hss}\box0}}
{\setbox0=\hbox{$\scriptstyle\rm H$}\hbox{\hbox to0pt
{\kern0.4\wd0\vrule height0.9\ht0\hss}\box0}}
{\setbox0=\hbox{$\scriptscriptstyle\rm H$}\hbox{\hbox to0pt
{\kern0.4\wd0\vrule height0.9\ht0\hss}\box0}}}}
\def\Ol{{\mathchoice
{\setbox0=\hbox{$\displaystyle\rm O$}\hbox{\hbox to0pt
{\kern0.4\wd0\vrule height0.9\ht0\hss}\box0}}
{\setbox0=\hbox{$\textstyle\rm O$}\hbox{\hbox to0pt
{\kern0.4\wd0\vrule height0.9\ht0\hss}\box0}}
{\setbox0=\hbox{$\scriptstyle\rm O$}\hbox{\hbox to0pt
{\kern0.4\wd0\vrule height0.9\ht0\hss}\box0}}
{\setbox0=\hbox{$\scriptscriptstyle\rm O$}\hbox{\hbox to0pt
{\kern0.4\wd0\vrule height0.9\ht0\hss}\box0}}}}
\newcommand{\ca}{\mathcal A}

\newcommand{\fc}{\mathfrak{c}}  \newcommand{\Fc}{\mathfrak{C}}

  \newcommand{\Ff}{\mathfrak{F}}
  
\newcommand{\fh}{\mathfrak{h}}  \newcommand{\Fh}{\mathfrak{H}}

  \newcommand{\Fl}{\mathfrak{L}}

\newcommand{\G}{\Gamma}
\newcommand{\eps}{\epsilon}

\DeclareMathOperator{\MCO}{\boldsymbol{\widehat{\mathsf{M}}}}

\DeclareMathOperator{\MCPW}{\rm Master\;\;Constraint\;\;Programme}

\title{Testing the\\ $\MCPW$\\ for Loop Quantum Gravity\\
III. SL(2,R) Models}
\author{B.
Dittrich\thanks{dittrich@aei.mpg.de, bdittrich@perimeterinstitute.ca},
T. Thiemann\thanks{thiemann@aei.mpg.de, tthiemann@perimeterinstitute.ca}\\
\\
Albert Einstein Institut, MPI f. Gravitationsphysik\\
Am M\"uhlenberg 1, 14476 Potsdam, Germany\\
\\
and\\
\\
Perimeter Institute for Theoretical Physics \\
31 Caroline Street North, Waterloo, ON N2L 2Y5, Canada}
\date{{\small Preprint AEI-2004-118}}
\begin{document}

\maketitle

\begin{abstract}
This is the third paper in our series of five in which we test the Master 
Constraint Programme for solving the Hamiltonian constraint in Loop 
Quantum Gravity. In this work we analyze models which, despite the 
fact that the phase space is finite dimensional, are much more complicated 
than in the second paper: These are systems with an $SL(2,\Rl)$ gauge 
symmetry and the complications arise because non -- compact 
semisimple Lie groups are not amenable (have no finite translation 
invariant measure). This leads to severe obstacles in the refined 
algebraic quantization programme (group averaging) and we see a trace 
of that in the fact that the spectrum of the Master Constraint does 
not contain the point zero. However, the minimum of the spectrum is of 
order $\hbar^2$ which can be interpreted as a normal ordering 
constant arising from first class constraints (while second class 
systems lead to $\hbar$ normal ordering constants). The physical Hilbert 
space can then be be obtained after subtracting this normal ordering 
correction.
 \end{abstract}

\newpage

\tableofcontents

\section{Introduction}
\label{s1}

We continue our test of the Master Constraint Programme 
\cite{7.0} for Loop Quantum Gravity (LQG) \cite{1.1,7.2,7.3}
which we started in the companion papers \cite{I,II} and will continue
in \cite{IV,V}. The Master Constraint Programme is a new idea to improve 
on 
the current situation with the Hamiltonian constraint operator for LQG
\cite{7.1}. In short, progress on the solution of the Hamiltonian 
constraint has been slow because of a technical reason: the Hamiltonian 
constraints 
themselves are not spatially diffeomorphism invariant. This means that one 
cannot first solve the spatial diffeomorphism constraints and then 
the Hamiltonian constraints because the latter do not preserve the 
space of solutions to the spatial diffeomorphism constraint
\cite{8.2}. On the other hand, the space of solutions  
to the spatial diffeomorphism constraint \cite{8.2} is relatively easy to 
construct starting from the spatially diffeomorphism invariant 
representations on which LQG is based \cite{7.4} which are therefore 
very natural to use and, moreover, essentially unique. Therefore one would 
really like to keep these structures. The Master Constraint 
Programme removes that technical obstacle by replacing the Hamiltonian 
constraints by a single Master Constraint which is a spatially 
diffeomorphism invariant integral of squares of the individual 
Hamiltonian constraints which encodes all the necessary information about 
the constraint surface and the associated invariants. See e.g. 
\cite{7.0,I} for a full discussion of 
these issues. Notice that the idea of squaring constraints is not new,
see e.g. \cite{2.1}, however, our concrete implementation is new and also 
the Direct Integral Decomposition (DID) method for solving them, see 
\cite{7.0,I} for all the details. 

The Master Constraint for four 
dimensional General Relativity will appear in \cite{8.1} but before we 
test its semiclassical limit, e.g. using the methods of \cite{8.3,8.5} 
and try to solve it by DID methods we want to test the programme in the  
series of papers \cite{I,II,IV,V}. 
In the previous papers we focussed on finite dimensional
systems of various degrees of complexity. 
In this article we will apply the Master Constraint Progamme to 
 constraint algebras which generate a non-abelean and non-compact gauge 
 group. In the first example we are concerned with the gauge group 
 $SO(2,1)$ and in the second example with the gauge group $SL(2,\Rl)$, 
 which is the double cover of $SO(2,1)$.

 We will see that both examples share the same problem -- the spectrum of 
 the Master Constraint Operator does not include the value zero. The 
 reason for this is the following: The value zero in the spectrum of the 
 Master Constraint Operator corresponds to the appearence of the trivial 
 representation in a Hilbert space decomposition of the given unitary 
 representation of the gauge group on the kinematical Hilbert space.

 Now, the groups $SO(2,1)$ and $SL(2,\Rl)$ (and all groups which have 
 these two groups as subgroups, e.g. symplectic groups and $SO(p,q)$ with 
 $p,q>1$ and $p+q>2$ ) are non-amenable groups, see \cite{9.2}. One 
 characteristic of non-amenable groups is, that the trivial representation 
 does not appear in a Hilbert space decomposition of the regular 
 representation into irreducible unitary subrepresentations. Since the 
 decomposition of the regular decomposition is often used to decompose 
 tensor products, it will often happen, that a given representation of a 
 non-amenable group does not include the trivial representation in its 
 Hilbert space decomposition. 

 Since the value zero is not included in the spectrum of the Master 
Constraint Operator $\MCO$ we will use a redefined operator 
$\MCO'=\MCO-\lambda_{\text{min}}$ as proposed in \cite{I}, where 
$\lambda_{\text{min}}$ is the 
 minimum of the spectrum. One can interprete this procedure as a quantum 
 correction (which is proportional to $\hbar^2$). Nevertheless the 
 redefinition of the Master Constraint Operator has to be treated 
 carefully, since it is not guaranteed that all relations between the 
 observables implied by the constraints are realized on the resulting 
 physical Hilbert space. This phenomenon will occur in the second example. 
 However we will show that it is possible to alter the Master Constraint 
 Operator again and to obtain a physical quantum theory which has the 
correct classical limit.

 In both examples we will use the representation theory of $SL(2,\Rl)$ and 
 its covering groups to find the spectra and the direct integral 
 decompositions with respect to the Master Constraint Operator. As we will 
 see, the diagonalization of the Master Constraint Operator is equivalent 
 to the diagonalization of the Casimir Operator of the given gauge group 
 representation, which in turn is equivalent to the decomposition of the 
 given gauge group representation into a direct sum and/or direct integral 
 of irreducible unitary representations.

 Furthermore, we will see that our examples exhibit the structure of a 
 dual pair, see \cite{howephys}. These are defined to be two subgroups in 
 a larger group, where one subgroup is the maximal commutant of the other 
 and vice versa. In our examples one subgroup is the group generated by 
 the observables and the other is the group generated by the constraints. 
 Now, given this structure of dual pairs, one can show that the 
 decomposition of the representation of the gauge group is equivalent to 
 the reduction of the representation of the observable algebra (on the 
 kinematical Hilbert space). This is explained in further detail in 
 appendix \ref{apposci}. In our examples this fact will help us to 
 determine the induced representation of the observable algebra on the 
 physical Hilbert space. Moreover we can determine in this way the induced 
 inner product on the physical Hilbert space, so that it will not be 
 necessary to perform all the steps of the direct Hilbert space 
 decomposition as explained in \cite{I} to find the physical inner 
 product. 

 We summarized the representation theory of the $sl(2,\Rl)$ algebra in 
 appendix \ref{appsl2r}. Appendix \ref{apposci} explains the theory of 
 oscillator representations for $sl(2,\Rl)$, which is havily used in the 
 two examples. It also contains a discussion how the representation theory 
 of dual pairs can be applied to our and similar examples.

\section{$SL(2,\Rl)$ Model with Non -- Compact Gauge Orbits}
\label{s5.2}

Here we consider the configuration space $\Rl^3$ with the three $so(2,1)$-generators as constraints:
\ba
L_i=\eps_{\,\, ij}^k x^j p_k, \q   \quad \{L_i,L_j\}=\eps_{\,\, ij}^k L_k
% L_1=-x^2p_3-x^3 p_2 \\
% L_2= x^3p_1+x^1p_3 \\
%L_3=x^1p_2-x^2 p_1
\ea
where $\eps_{\,\, ij}^k=g^{km}\eps_{mij}$,\,\, $\eps_{ijk}$ is totally antisymmetric with $\eps_{123}=1$ and $g^{ik}$ is the inverse of the metric $g_{ik}=\text{diag}(+,+,-)$. Indices are raised and lowered with $g^{ik}$ resp. $g_{ik}$ and we sum over repeated indices.

The gauge group $SO(n,1)$ was previously discussed in \cite{gomberoff}, where group averaging was used to construct the physical Hilbert space. We will compare the results of \cite{gomberoff} and the results obtained here at the end of the section.

The observable algebra of the system above is generated by
\ba \label{obsmink}
d=x^ip_i \quad \q e^+=x^ix_i \quad \q e^-=p^ip_i  \q.
\ea 
This set of observables exhibits the commutation relations of the generators of the $sl(2,\Rl)$-algebra (which coincides with $so(2,1)$):
\ba
\{d,e^\pm\}=\mp 2 e^\pm   \q \quad \{e^+,e^-\}=4d \q.
\ea
 We have the identity
\be \label{classidmink}
d^2-e^+ e^-=L_iL^i
\ee
between the Casimirs of the constraint and observable algebra. 

\subsection{Quantization}

We start with the auxilary Hilbert space $\Fl^2(\Rl^3)$ of square integrable functions of the coordinates. The momentum operators are $\hat p_j=-i(\hbar)\partial_j$ and the $\hat x^j$ act as multiplication operators. There arises no factor ordering ambiguity for the quantization of the constraints, but to ensure a closed observable algebra, we have to choose:
\ba \label{sl2rrepmink}
\hat d &=& \tfrac{1}{2}(\hat x^i\hat p_i+\hat p_i \hat x^i)=\hat x^i  \hat p_i - \tfrac{3}{2}i \hbar \nn \\
\hat e^+ &=& \hat x^i \hat x_i \quad \q \q \hat e^-=\hat p^i \hat p_i 
\ea
The commutators between constraints and between observables are then obtained by replacing the Poisson bracket with $\tfrac{1}{i\hbar}\big[\cdot,\cdot\big]$.

The identity (\ref{classidmink}) is altered to be:
\ba \label{quidmink}
\hat d^2-\tfrac{1}{2}(\hat e^+ \hat e^-+ \hat e^- \hat e^+)-\tfrac{3}{4}\hbar^2 = \hat L_i \hat L^i \q.
\ea

(From now on we will skip the hats and set $\hbar$ to 1.)

For the implementation of the Master Constraint Programme we have to construct the spectral resolution of the $\MCO$
\be
\MCO:=L_1^2+L_2^2+L_3^2=L^i L_i + 2L_3^2=  d^2-\tfrac{1}{2}( e^+  e^-+  e^-  e^+)-\tfrac{3}{4}+2L_3^2 \q.
\ee 
To this end we will use the following strategy:  
 The operators $\MCO$ and $L_3$ commute, so we can diagonalize them 
 simultaneously. The diagonalization of $L_3$ is easy to achieve, its 
 spectrum being purely discrete, namely $\text{spec}(L_3)=\Zl$. Now we can 
 diagonalize $\MCO$ on each eigenspace of $L_3$ seperately. On these 
 eigenspaces the diagonalization of $\MCO$ is equivalent to the 
 diagonalization of the $so(2,1)$-Casimir $L_iL^i$ and because of identity 
 (\ref{quidmink}) equivalent to the diagonalization of the 
 $sl(2,\Rl)$-Casimir $\Fc=-\tfrac{1}{4}(d^2-\tfrac{1}{2}( e^+  e^-+  e^-  
 e^+))$. As we will show below the $sl(2,\Rl)$-representation given by 
(\ref{sl2rrepmink}) is a tensor product of three representations, which are known as oscillator and contragredient oscillator representations. To obtain the spectral resolution of the Casimir $\Fc$, we will reduce this tensor product into its irreducible components.

\subsection{The Oscillator Representations and its Reduction}

For the reduction process it will be very convenient to work with the following basis of the $sl(2,\Rl)$-algebra:
\begin{alignat}{2} \label{osci1mink}
h &=\tfrac{1}{4}(e^++e^-)&&=\tfrac{1}{2}(a_1^\dagger a_1+a_2^\dagger a_2-a_3^\dagger a_3 +\tfrac{1}{2}) \nn \\
n^+ &=-\tfrac{1}{2}(i \,d-\tfrac{1}{2}(e^+-e^-))&&=\tfrac{1}{2}(a_1^\dagger a_1^\dagger +a_2^\dagger a_2^\dagger -a_3 a_3) \nn \\
n^- &=\tfrac{1}{2}(i \,d+\tfrac{1}{2}(e^+-e^-))&&=\tfrac{1}{2}(a_1 a_1 +a_2 a_2 -a_3^\dagger a_3^\dagger) \nn \\
 \Fc &= \tfrac{1}{4}(d^2-\tfrac{1}{2}( e^+  e^-+  e^-  e^+))&&=-h^2+\tfrac{1}{2}(n^+ n^-+n^-n^+) \q\q,
\end{alignat}
with commutation and adjointness relations
\ba
\big[h,n^\pm \big]=\pm n^\pm \q \q  \big[n^+,n^-\big]=-2h \q \q (n^+)^\dagger=n^- \q.
\ea
Here we introduced the anihilation and creation operators 
\be
a_i=\tfrac{1}{\sqrt{2}}(x^i+ip_i) \q\text{and} \q a_i^\dagger=\tfrac{1}{\sqrt{2}}(x^i-ip_i) \q .
\ee
Now it is easy to see that this representation is a tensor product of the following three $sl(2,\Rl)$-representations:
\begin{alignat}{2} \label{factors}
&h_i=\tfrac{1}{2}(a_i^\dagger a_i +\tfrac{1}{2})\q \text{for}\,\,i=1,2 \q \text{and}\q & &h_3=-\tfrac{1}{2}(a_3^\dagger a_3+\tfrac{1}{2}) \nn \\
&n^+_i=\tfrac{1}{2}a_i^\dagger a_i^\dagger  &  &n^+_3=-\tfrac{1}{2}a_3 a_3 \nn \\
&n^-_i=\tfrac{1}{2}a_i a_i  & &n^-_3=-\tfrac{1}{2}a_3^\dagger a_3^\dagger
\end{alignat}
These representations are known as oscillator representation $\omega$ (for $i=1,2$) and contragredient oscillator representation $\omega^*$ (for $i=3$), see \cite{howe} and \ref{apposci}, where these representations are explained.

 The oscillator representation is the sum of two irreducible representations, which are the representations $D(1/2)$ and $D(3/2)$ from the positive discrete series of the double cover of $Sl(2,\Rl)$ (corresponding to even and odd number Fock states). Similarly $\omega^* \simeq D^*(1/2)\oplus D^*(3/2)$, where $D^*(1/2)$ and $D^*(3/2)$ are from the negative discrete series.

As mentioned before we will reduce this tensor product to its irreducible subrepresentations in order to obtain the spectrum of the Casimir (and with it the spectrum of the Master Constraint Operator). In appendix \ref{apposci} one can find the general strategy and some formulas to reduce such tensor products. Furthermore appendix \ref{appsl2r} reviews the $sl(2,\Rl)$-representations, which will appear below.  

To begin the reduction of $\omega\otimes\omega\otimes\omega^*$ we will reduce $\omega\otimes\omega$. This we can achieve by utilizising the observable $L_3$. It commutes with the $sl(2,\Rl)$-algebra (\ref{osci1mink}), therefore according to Schur's Lemma its eigenspaces are left invariant by the $sl(2,\Rl)$-algebra, i.e. its eigenspaces are subrepresentations of $sl(2,\Rl)$. 

To diagonalize $L_3$ and reduce the tensor product $\omega \otimes \omega$ we will employ the ``polarized'' anihilation and creation operators
\ba
A_{\pm}=\frac{1}{\sqrt{2}}(a_1 \mp i a_2) \quad \quad A_{\pm}^\dagger=\frac{1}{\sqrt{2}}(a_1 \pm i a_2).
\ea   
With the help of these, we can write
\ba
h &=&\tfrac{1}{2}(A_+^\dagger A_++A_-^\dagger A_--a_3^\dagger a_3 +\tfrac{1}{2}) \nn \\
n^+ &=&A_+^\dagger A_-^\dagger -\tfrac{1}{2}a_3 a_3 \nn \\
n^-&=&A_+ A_- - \tfrac{1}{2}a_3^\dagger a_3^\dagger  \label{rep3} \\
L_3&=& A_+^\dagger A_+ - A_-^\dagger A_- \q.
\ea
In the following we will denote by $|k_+,k_-,k_3>$ Fock states with respect to $A_+^\dagger$, $A_-^\dagger$ and $a_3^\dagger$. The operator $L_3$ acts on them diagonally. Its eigenspaces $V(\pm j)$ corresponding to the eigenvalue $\pm j,\,j\in\Nl$ are generated by $\{|j,0,k_3>,k_3 \in \Nl\}$ and $\{|0,j,k_3>,k_3\in \Nl\}$ respectively, i.e. $V(j)$ is (the closure of) the linear span of the $|j,0,k_3>$'s resp. $|0,j,k_3>$'s and all vectors are obtained by applying repeatedly $n^+$ to them.  These eigenspaces are invariant subspaces of the representation (\ref{rep3}). This representation restricted to $V(j)$ is still a tensor product representation, namely the representation $D(|j|+1)\otimes\omega^*$.
Its factors are given by 
\begin{alignat}{2}
&h_{12}|_{V(j)}=\tfrac{1}{2}(A_+^\dagger A_++A_-^\dagger A_- +1)|_{V(j)}\q\q &\text{and}\q \q  &h_3|_{V(j)}=-\tfrac{1}{2} (a_3^\dagger a_3 +\tfrac{1}{2})|_{V(j)} \nn \\
&n^+_{12}|_{V(j)}=A_+^\dagger A_-^\dagger|_{V(j)}\q\q &\text{and}\q \q &n^+_{3}|_{V(j)}=- \tfrac{1}{2}a_3 a_3 \nn \\
&n^-_{12}|_{V(j)} =A_+ A_-|_{V(j)}\q\q &\text{and}\q  \q &n^-_{3}|_{V(j)}= - \tfrac{1}{2}a_3^\dagger a_3^\dagger \q .
\end{alignat}
Since $h_{12}$ has a smallest eigenvalue $\tfrac{1}{2}(|j|+1)$ on the subspace $V(j)$, this subspace carries a $D(|j|+1)$-representation from the positive discrete series (of $SL(2,\Rl)$) with lowest weight $\tfrac{1}{2}(|j|+1)$ (see \ref{appsl2r}).
 % It is characterized by an $h$-spectrum $\text{spec}(h)=\{\tfrac{1}{2}(|j|+1+k),k \in \Nl\}$ and a Casimir  $\Fc(D(|j|+1))=(-\tfrac{1}{4}(|j|+1)^2+\tfrac{1}{2}(|j|+1))\text{Id}$. 

So far we have achieved the reduction $\omega \otimes \omega \otimes \omega^*\simeq (D(1)\oplus \sum_{j=0}^\infty 2 D(j+1))\otimes \omega^*$.
To reduce the representation (\ref{rep3}) completely, we have to consider tensor products of the form $D(|j|+1)\otimes D^*(1/2)$ and $D(|j|+1)\otimes D^*(3/2)$. We take this reduction from \cite{howe}, see also \ref{apposci} and \ref{appsl2r} for a description of the $sl(2,\Rl)$-representations, appearing below:

For $j$ even, we have
\ba
D(|j|+1)\otimes D^*(1/2)\simeq \int_{\tfrac{1}{2}}^{\tfrac{1}{2}+i\infty} P(t,1/4) d\mu(t)\oplus \sum_l  D(|j|+1/2-2l) \nn \\  \text{with}  \; 0 \leq 2l<|j|-1/2,\,l \in \Nl \nn \\
\ea
and for $j$ odd we get
\ba
D(|j|+1)\otimes D^*(1/2)\simeq \int_{\tfrac{1}{2}}^{\tfrac{1}{2}+i\infty} P(t,-1/4) d\mu(t)\oplus \sum_l  D(|j|+1/2-2l) \nn \\  \text{with}  \; 0 \leq 2l<|j|-1/2,\,l \in \Nl \q. \nn \\
\ea 
In particular, we have for $j=0$
\be
D(1)\otimes D^*(1/2)\simeq \int_{\tfrac{1}{2}}^{\tfrac{1}{2}+i\infty} P(t,1/4) d\mu(t) \q.
\ee
The remainig tensor products are
\ba
D(1)\otimes D^*(3/2) &\simeq &\int_{\tfrac{1}{2}}^{\tfrac{1}{2}+i\infty} P(t,-1/4) d\mu(t) \nn \\
D(|j|+1)\otimes D^*(3/2) &\simeq& D(|j|)\otimes D^*(1/2) \q \text {for}\;j>0 \q .
\ea
$P(t,\eps),\eps=\tfrac{1}{4},-\tfrac{1}{4}$ is the principal series (of the metaplectic group, i.e. the double cover of $SL(2,\Rl)$) characterized by an $h$-spectrum $\text{spec}(h)=\{\eps+z,z\in \Zl\}$ and a Casimir $\Fc(P(t,\eps))=t(1-t)\text{Id}$. The measure $d\mu(t)$ is the Plancherel measure on the unitary dual of the metaplectic goup.

The representations $D(l+1/2),l \in \Nl-\{0\}$ are positive discrete series representations of the metaplectic group. The $h$-spectrum in these representations is given by $\{\tfrac{1}{2} (l+1/2)+n,n\in\Nl\}$ and the Casimir by $\Fc(D(l+1/2))=-\tfrac{1}{4} (l+1/2)^2+\tfrac{1}{2} (l+1/2)$.

The spectrum of the Casimir $\Fc$ is non-degenerate on each tensor product $D(k)\otimes D^*(l),\,l=\tfrac{1}{2},\tfrac{3}{2}$, i.e. the Casimir discriminates the irreducible representations, which appear in this tensor product and the irreducible representations have multiplicity one. The spectrum of $h$ is nondegenerate in each irreducible representation of the metaplectic group . This implies that we can find a (generalized) basis $|j,\eps,\fc,\fh>$, which is labeled by the $L_3$-eigenvalue $j$, the values $\eps=\tfrac{1}{4},-\tfrac{1}{4}$, the Casimir eigenvalue $\fc$ and the $h$-eigenvalue $\fh$.       

Summarizing, we have for the (highly degenerate) spectrum of the Casimir $\Fc=-h^2+\tfrac{1}{2} (n^+ n^-+n^-n^+)$ on $\Fl^2(\Rl^3)$:
\be
\text{spec}(\Fc)=\{\tfrac{1}{4} +x^2,x\in \Rl,x \geq 0\} \cup \{ -\tfrac{1}{4} (l+1/2)^2+ \tfrac{1}{2}(l+1/2),l \in \Nl-\{0\}\} \q .
\ee
The continuous part of the spectrum originates from the principal series $P(t,1/4)$ and $P(t,-1/4)$ and the discrete part from those positive discrete series representations $D(l)$, which appear in the decompositions above.
This results in the following expression for the spectrum of the $so(2,1)$-Casimir $L^iL_i=(4\Fc-\tfrac{3}{4})$:
\be
\text{spec}( L^iL_i)=\{\tfrac{1}{4} +s^2,s\in \Rl\,s\geq 0\} \cup \{-q^2+q,q \in \Nl-\{0\}\} \q .
\ee
As explained in appendix \ref{appsl2r} these values correspond to the principal series $P(\tfrac{1}{2}+s,0)$ of $SO(2,1)$ and the positive or negative discrete series $D(2q)$ resp. $D^*(2q)$ for $q\in \Nl-\{0\}$. In the $SO(2,1)$-principal series the spectrum of $L_3$ is given by $\text{spec}(L_3)=\{\Zl\}$. In the discrete series $D(2q)$ we have $\text{spec}(L_3)=\{q+n,n\in\Nl\}$ and in $D^*(q)$\; $\text{spec}(L_3)=\{-q-n,n\in\Nl\}$.

\subsection{The Physical Hilbert space}

Now we can determine the spectrum of the Master Constraint Operator $\MCO=L_iL^i+2L_3^2$ on the $j$-eigenspaces $V(j)$ of $L_3$. From the principal series we get the continous part of the spectrum
\be
\text{spec}_{\text{cont}}(\MCO|_{V(j)})=\{\tfrac{1}{4}+s^2+2j^2,s \in \Rl,s\geq0\}
\ee
and from the discrete series the discrete part (for $j\geq 1$, since there is no discrete part for $j=0$)
\be
\text{spec}_{\text{discr}}(\MCO|_{V(j)})=\{(-q^2+q)+2j^2,q \in \Nl,q\leq |j|\}\geq 2 \,\, .
\ee
(The inequality $q\leq |j|$ follows from the fact, that in a representation $D(2q)$ or $D^*(2q)$ we have $|j|\geq q$ for the $L_3$-eigenvalues $j$.)

 One can see immedeatily, that zero is not included in the spectrum of the 
 Master Constraint, the lowest generalized eigenvalue being 
$\tfrac{1}{4}$. Therefore we alter the 
Master Constraint to $\MCO'=\MCO-\tfrac{1}{4}(\hbar^2)$ where appropriate 
powers of $\hbar$ have been restored. The generalized 
null eigenspace of $\MCO'$ is given by the linear span of all states $|j=0,\eps,\fc=\tfrac{1}{4},\fh>$. The spectral measure of $\MCO'$ induces a scalar product on this space, which can then be completed to a Hilbert space. In particular with this scalar product one can normalize the states $|j=0,\eps,\fc=\tfrac{1}{4},\fh>$, obtaining an ortho-normal basis $||\eps,\fh>>$. 

This Hilbert space has to carry a unitary representation of the metaplectic group. Actually, it carries a sum of two irreducible representations $P(t=1/2,1/4)$ and $P(t=1/2,-1/4)$, corresponding to the labels $\eps=1/4$ and $\eps=-1/4$ of the basis \{$||\eps,\fh>>\}$. States in these representations are distinguished by the transformation under the reflection $R_3:x_3\mapsto -x_3$, which is a group element of $O(2,1)$. As an operator on $\Fl^2(\Rl^3)$ it acts as:
\be
\hat R_3:\psi(x_1,x_2,x_3) \mapsto \psi(x_1,x_2,-x_3) \q .
\ee
$\hat R_3$ acts on states with $\eps=\tfrac{1}{4}$ as the identity operator (since these states are linear combinations of even number Fock states with respect to $a_3^\dagger$) and on states with $\eps=-\tfrac{1}{4}$ by multiplying them with $(-1)$ (since these states are linear combinations of odd number Fock states).   It seems natural, to exclude the states with nontrivial behaviour under this reflection. This leaves us with the unitary irreducible representation $P(t=1/2,1/4)$. As explained in \ref{appsl2r} the action of the observable algebra $sl(2,\Rl)$ on the states $||\fh>>:=||1/4,\fh>>$ is determined (up to a phase, which can be fixed by adjusting the phases of the states $||\fh>>$) by this representation to be:
\ba
 h||\fh>>&=&\fh||\fh>> \q \q (\fh \in \{\tfrac{1}{4}+\Zl\}) \nn \\
 n^+||\fh>>&=&(\fh+\tfrac{1}{2})||\fh+1>> \nn \\
 n^-||\fh>>&=&(\fh-\tfrac{1}{2}) ||\fh-1>> \q .
\ea
This gives for matrix elements of the observables $d$ and $e^\pm$ (see \ref{sl2rrepmink})
\ba
<<\fh'|d|\fh>>&=&i(\fh+\tfrac{1}{2})\delta_{\fh',\fh+1}-i(\fh-\tfrac{1}{2})\delta_{\fh',\fh-1}  \\   
<<\fh'|e^\pm|\fh>>&=&2 \fh\,\delta_{\fh',\fh} \pm (\fh+\tfrac{1}{2})\delta_{\fh',\fh+1} \pm (\fh-\tfrac{1}{2})\delta_{\fh',\fh-1} \q.
\ea

 The operators $e^+=\hat{x}^i \hat{x}_i$ and $e^-=\hat{p}^i \hat{p}_i$ are 
 indefinite operators, i.e. their spectra include positive and negative 
numbers.

 To sum up, we obtained a physical Hilbert space, which carries an 
 irreducible unitary representation of the observable algebra. In contrast 
 to these results the group averaging procedure in \cite{gomberoff} leads 
 to (two) superselection sectors and therefore to a reducible 
 representation of the observables. These sectors are functions with 
 compact support inside the light cone and functions with compact support 
 outside the light cone. Hence the observable $e^+=\hat{x}^i \hat{x}_i$ is 
 either strictly positive or strictly negative definite on theses 
 superselection sectors. From that point of view our physical 
Hilbert space is preferred because physically $e^+$ should be 
indefinite. However, as mentioned in \cite{gomberoff} the 
appearence of superselection sectors may depend on the choice of the 
 domain $\Phi$, on which the group averaging procedure has to be defined
and thus other choices of $\Phi$ may not suffer from this superselection
problem. We see, at least in this example, that the DID method 
outlined in \cite{I} with the prescription given there gives a more 
natural and unique result.
However, as the given system lacks a realistic interpretation anyway,
this difference may just be an artefact of a pathological model.

\section{Model with Two Hamiltonian Constraints and Non -- Compact
Gauge Orbits}
\label{s5.3}

\subsection{Introduction of the Model}

Here we consider a reparametrization invariant model introduced by Montesinos, Rovelli and T.T. in \cite{rov}. It has an $Sl(2,\Rl)$ gauge symmetry and a global $O(2,2)$ symmetry and has attracted interest because its constraint structure is in some sense similar to the constraint structure found in general relativity. Further work on this model has appeared in \cite{louko,louko2,trunk,gamb} and references therein.

We will shortly summarize the classical (canonical) theory (see \cite{rov} for an extended discussion). The configuration space is $\Rl^4$ parametrized by coordinates $(u_1,u_2)$ and $(v_1,v_2)$ and the canonically conjugated momenta are $(p_1,p_2)$ and $(\pi_1,\pi_2)$. The system is a totally constrained (first class) system. The constraints form a realization of an $sl(2,\Rl)$-algebra:    
\ba
H_1=\tfrac{1}{2}(\vec{p}^2-\vec{v}^2) \hspace{1cm}
H_2=\tfrac{1}{2}(\vec{\pi}^2-\vec{u}^2) \hspace{1cm}
D=\vec{u}\cdot\vec{p}-\vec{v}\cdot\vec{\pi} 
\ea
\be
\{H_1,H_2\}=D \q \q \{H_1,D\}=-2H_1 \q\q \{H_2,D\}=2H_2
\ee
The canonical Hamiltonian governing the time evolution (which is pure gauge) is $H=N\,H_1+M\,H_2+\lambda\,D$ where $N,M$
and $\lambda$ are Lagrange multipliers. Since $H_1$ and $H_2$ are quadratic in the momenta and their Poisson bracket gives a constraint which is linear in the momenta, one could say that this model has an analogy with general relativity. There, one has Hamiltonian constraints $H(x)$ quadratic in the momenta and diffeomorphism constraints $D(x)$ linear in the momenta which have the Poisson structure $\{H(x),H(y)\}\sim \delta(x-y)D(x)$ and $\{H(x),D(y)\}\sim \delta(x-y) H(x)$.

However, one can make the following canonical transformation to new canonical coordinates $(U_i,V_i,P_i,\Pi_i),\,i=1,2$ that transforms the constraint into phase space functions which are linear in the momenta:
\begin{xalignat}{2}
&u_i=\tfrac{1}{\sqrt{2}}(U_i+\Pi_i) & & v_i=\tfrac{1}{\sqrt{2}}(V_i+P_i) \nn \\
&p_i=\tfrac{1}{\sqrt{2}}(-V_i+P_i)  & & \pi_i=\tfrac{1}{\sqrt{2}}(-U_i+\Pi_i)
\end{xalignat}
\ba
H_1=-P_1 V_1-P_2 V_2   \q\q H_2=-U_1 \Pi_1 -U_2 \Pi_2  \q\q
D=P_1U_1+P_2U_2-V_1\Pi_1-V_2\Pi_2 \q .
\ea
These coordinates have the advantage, that the constraints act on the configuration variables $(U_1,V_1)$ and $(U_2,V_2)$ in the defining two-dimensional representation of $sl(2,\Rl)$ (i.e. by matrix multiplication). 

For reasons that will become clear later, it is easier for us to stick to the old coordinates $(u_i,v_i,p_i,\pi_i)$.

Now we will list the Dirac observables of this system. They reflect the global $O(2,2)$-symmetry of this model and are given by (see \cite{rov})
\begin{xalignat}{2}
& O_{12}=u_1 p_2-p_1 u_2 &&    O_{23}=u_2 v_1-p_2 \pi_1 \nonumber \\
& O_{13}=u_1 v_1-p_1 \pi_1 &&  O_{24}=u_2 v_2-p_2 \pi_2 \nonumber \\
& O_{14}=u_1 v_2-p_1 \pi_2 &&  O_{34}=\pi_1 v_2-v_1 \pi_2
\end{xalignat}
They constitute the Lie algebra $so(2,2)$ which is isomorphic to $so(2,1)\times so(2,1)$. A basis adapted to the $so(2,1)\times so(2,1)$-structure is (see \cite{trunk})
\begin{xalignat}{2}  \label{obssl}
& Q_1=\tfrac{1}{2}(O_{23}+O_{14}) & & P_1=\tfrac{1}{2}(O_{23}-O_{14}) \nonumber \\
& Q_2=\tfrac{1}{2}(-O_{13}+O_{24}) & & P_2=\tfrac{1}{2}(-O_{13}-O_{24}) \nonumber \\     
& Q_3=\tfrac{1}{2}(O_{12}-O_{34}) & & P_3=\tfrac{1}{2}(O_{12}+O_{34}) 
\end{xalignat}
The Poisson brackets between these observables are
\ba
\{Q_i,Q_j\}=\eps_{ij}^{\,\,\,\,\,k}Q_k \q \q \{P_i,P_j\}=\eps_{ij}^{\,\,\,\,\,k}P_k \q \q \{Q_i,P_j\}=0
%&[Q_1,Q_2]=-iQ_3 \quad &[Q_2,Q_3]=iQ_1 \quad &[Q_3,Q_1]=iQ_2 \nonumber \\
%&[P_1,P_2]=-iP_3 \quad &[P_2,P_3]=iP_1 \quad &[P_3,P_1]=iP_2 \nonumber \\
%&[Q_i,P_j]=0
\ea
where $\eps_{ij}^{\,\,\,\,\,k}=g^{lk} \eps_{ijk}$, with $g^{lk}$ being the inverse of the metric $g_{lk}=\text{diag}(+1,+1,-1)$. The Levi-Civita symbol $\eps_{ijk}$ is totally antisymmetric with $\eps_{123}=1$ and we sum over repeated indices. 
Lateron the (ladder) operators $Q_\pm:=\tfrac{1}{\sqrt{2}}(Q_1\pm iQ_2)$ and $P_\pm:=\tfrac{1}{\sqrt{2}}(P_1\pm iP_2)$ will be usefull.

One can find the following identities between observables and constraints (see \cite{trunk}):
\ba \label{clid}
&& Q_1^2+Q_2^2-Q_3^2 = P_1^2+P_2^2-P_3^2=\tfrac{1}{4}(D^2+4H_1 H_2) \\
&& 4 Q_3 P_3 = (\vec{u}^2-\vec{v}^2)(H_1+H_2)-(\vec{u}\cdot\vec{p}+\vec{v}\cdot\vec{\pi})D+(\vec{u}^2+\vec{v}^2)(H_1-H_2)
\ea
 They imply that on the constraint hypersurface we have $Q_i=0 \,\, 
 \forall i$ or $P_i=0 \,\, \forall i$. %All observables vanish on a 
 certain submanifold of the phase space $\Rl^8$. Notice that the 
constraint hypersurface consists of the disjoint union of the following 
five varieties: $\{Q_j=P_j=0,\;j=1,2,3\},
\;\{\pm Q_3>0,\;P_3=0\},\;\{Q_3=0,\;\pm P_3>0\}$.

\subsection{Quantization}

For the quantization we will follow \cite{rov} and choose the coordinate representation where the momentum operators act as derivative operators and the configuration operators as multiplication operators on the Hilbert space $\Fl^2(\Rl^4)$ of square integrable functions $\psi(\vec{u},\vec{v})$:
\begin{xalignat}{2}
&\hat{\vec{p}}\psi(\vec{u},\vec{v})=-i\hbar \vec{\nabla}_u\psi(\vec{u},\vec{v}) &&  \hat{\vec{\pi}}\psi(\vec{u},\vec{v})=-i\hbar \vec{\nabla}_v \psi(\vec{u},\vec{v}) \nn \\
& \hat{u}_i\psi(\vec{u},\vec{v}) =u_i\psi(\vec{u},\vec{v}) &&
\hat{v}_i\psi(\vec{u},\vec{v}) =v_i\psi(\vec{u},\vec{v}) \q .
\end{xalignat}
In the following we will skip the hats and set $\hbar=1$.

For the constraint algebra to close we have to quantize the constraints in the following way:
\begin{xalignat}{3} \label{qsl}
&H_1 =-\tfrac{1}{2}(\Delta_u +\vec{v}^2) &&
H_2 = -\tfrac{1}{2}( \Delta_v +\vec{u}^2 )   &&
D =-i(\vec{u}\cdot\vec{\nabla}_u-\vec{v}\cdot\vec{\nabla}_v) \nn \\
& \big[H_1,H_2\big]=iD  &&   \big[D,H_1\big]=2iH_1 && \big[D,H_2\big]=-2iH_2 \q .
\end{xalignat}

There arises no factor ordering ambiguity for the quantization of the observable algebra. The algebraic properties are preserved in the quantization process, i.e. Poisson brackets between observables $O_{ij}$ are simply replaced by $-i\big[\cdot,\cdot\big]$. 

We introduce a more convenient basis for the constraints:
\ba \label{qcon}
H_+=H_1+H_2 \q\q
H_-=H_1-H_2 \q\q
D=D                \q.
\ea
$H_-$ is just the sum and difference of Hamiltonians for one-dimensional harmonic oscillators. (It is the generator of the compact subgroup $SO(2)$ of $Sl(2, \Rl)$ and has discrete spectrum in $ \Zl$).
The commutation relations are now:
\ba
\big[H_-,D\big]=-2iH_+ \q \q
\big[H_+,D\big]=-2iH_- \q \q
\big[H_+,H_-\big]=-2iD   \q.  
\ea
The operator
\be
\Fc=\tfrac{1}{4}(D^2+H_+^2-H_-^2)
\ee
commutes with all three constraints (\ref{qcon}), since it is the (quadratic) Casimir operator for $sl(2,r)$ (see Appendix \ref{appsl2r}). According to Schur's lemma, it acts as a constant on the irreducible subspaces of the $sl(2,\Rl)$ representation given by (\ref{qcon}).   

The quantum analogs of the classical identities (\ref{clid}) are
\ba \label{quid}
&&Q_1^2+Q_2^2-Q_3^2=P_1^2+P_2^2-P_3^2=\tfrac{1}{4}(D^2+H_+^2-H_-^2)=\Fc \\
&&4 Q_3 P_3 =(\vec{u}^2-\vec{v}^2)(H_+)-(\vec{u}\cdot\vec{p}+\vec{v}\cdot\vec{\pi})D+(\vec{u}^2+\vec{v}^2)(H_-) \q .
\ea

\subsection{The Oscillator Representation}

We are interested in the spectral decomposition of the Master Constraint Operator, which we define as
\be
\MCO=D^2+H_+^2+H_-^2=4\Fc+ 2 H_-^2.
\ee
The Master Constraint Operator is the sum of (a multiple of) the Casimir operator and $H_-$, which commutes with the Casimir. Therefore we can diagonalize these two operators simultaneously, obtaining a diagonalization of $\MCO$. We can achieve a diagonalization of the Casimir by looking for the irreducible subspaces of the $sl(2,\Rl)$-representation given by (\ref{qcon}), since the Casimir acts as a multiple of the identity operator on these subspaces. Hence we will attempt do determine the representation given by (\ref{qcon}).     

By introducing creation and annihilation operators
\begin{xalignat}{2}
& a_i=\frac{1}{\sqrt{2}}(u_i+\partial_{u_i}) && a_i^\dagger=\frac{1}{\sqrt{2}}(u_i-\partial_{u_i}) \\
& b_i=\frac{1}{\sqrt{2}}(v_i+\partial_{v_i}) && b_i^\dagger=\frac{1}{\sqrt{2}}(v_i-\partial_{v_i}) 
\end{xalignat}
we can rewrite the constraints as
\ba
H_- &=&\sum_{i=1,2}(a_i^\dagger a_i - b_i^\dagger b_i )\\
H_+ &=&-\frac{1}{2}\sum_{i=1,2}(a_i^2 + (a_i^\dagger)^2+ b_i^2+(b_i^\dagger)^2) \\
D &=& \frac{i}{2}\sum_{i=1,2}(-a_i^2 + (a_i^\dagger)^2+ b_i^2-(b_i^\dagger)^2)   \q .
\ea

This $sl(2,\Rl)$ representation is a tensor product of the following four representations (with $i \in \{1,2\}$):
\begin{xalignat}{2} \label{uivi1}
(h_-)_{u_i}&=a_i^\dagger a_i+\frac{1}{2} & (h_-)_{v_i} &= -b_i^\dagger b_i-\frac{1}{2} \nn \\
(h_+)_{u_i} &= -\frac{1}{2}((a_i^\dagger)^2+a_i^2) & (h_+)_{v_i} &= -\frac{1}{2}((b_i^\dagger)^2+b_i^2) \nn \\
d_{u_i} &= \frac{i}{2}((a_i^\dagger)^2-a_i^2) & d_{v_i} &= \frac{i}{2}(-(b_i^\dagger)^2+b_i^2)  
\end{xalignat}
The $u_i$-representations are known as oscillator representations $\omega$ and the $v_i$-representations as contragredient oscilalator representations $\omega^*$, see appendix \ref{apposci} for a discussion of these representations. As is also explained there these representation are reducible into two irreducible representations $D(1/2)$ and $D(3/2)$ for the oscillator representation $\omega$ and $D^*(1/2)$ and $D^*(3/2)$ for the contragredient oscillator representation $\omega^*$. The representation $D(1/2)$ respectively $D^*(1/2)$ acts on the space of even number Fock states, whereas $D(3/2)$ respectively $D^*(3/2)$ acts on the space of uneven Fock states. The representations $D(1/2)$ and $D(3/2)$ are members of the positive discrete series (of the two-fold covering group of $Sl(2,\Rl)$), $D^*(1/2)$ and $D^*(3/2)$ are members of the negative discrete series. (We have listed all $sl(2,\Rl)$-representations in appendix \ref{appsl2r}.) 
%As one can easily see, the space of even number Fock states and the space of uneven number Fock space is left invariant by the $sl(2,\Rl)$ generators. Therefore the oscillator rep is reducible into two subspaces, which are actually irreducible. The two irreps are $D(1/2)$ and $D(3/2)$ from the (positive) discrete series (of the two-fold covering of $Sl(2,\Rl)$).
% The contragredient oscillator representation is the direct sum of the irreps $D^*(1/2)$ and $D^*(3/2)$ from the negative discrete series.

Our aim is to reduce the tensor product $\omega \otimes \omega \otimes \omega^* \otimes \omega^*$ into its irreducible components. The isotypical component with respect to the trivial representation would correspond to the physical Hilbert space. 
To begin with we consider the tensor product $\omega \otimes \omega$. The discussion for $\omega^* \otimes \omega^*$ is analogous.

%Using the representation theory of $sl(2,\Rl)$ (see Appendix \ref{}), we find that
%\ba
%\omega \otimes \omega &=& (D(1/2) \oplus D(3/2))\otimes  (D(1/2) \oplus D(3/2)) \nonumber \\
%&=& D(1)\oplus \sum_{j=2}^{\infty} 2 D(j)
%\ea

To this end we utilize the observable $O_{12}$ (and $O_{34}$ for the tensor product $\omega^* \otimes \omega^*$). Since $O_{12}$ commutes with the $sl(2,\Rl)$-generators the eigenspaces of $O_{12}$ are $sl(2,\Rl)$-invariant. The observable $O_{12}$ is diagonal in the ``polarized'' Fock basis, which is defined as the Fock basis with respect to the new creation and annihilation operators
\be
A_{\pm}=\frac{1}{\sqrt{2}}(a_1 \mp i a_2) \quad \quad A_{\pm}^\dagger=\frac{1}{\sqrt{2}}(a_1^\dagger \pm i a_2^\dagger).
\ee 
The ``polarized'' creation and annihilation operators for the $v$-coordiantes are
\be
B_{\pm}=\frac{1}{\sqrt{2}}(b_1 \mp i b_2) \quad \quad B_{\pm}^\dagger=\frac{1}{\sqrt{2}}(b_1^\dagger \pm i b_2^\dagger).
\ee
With help of these operators we can write the $sl(2,\Rl)$-generators for the $\omega\otimes\omega$ representation and for the $\omega^*\otimes\omega^*$ representation as 
\begin{xalignat}{2} \label{firstred}
h_{-A}&=A_+^\dagger A_+ +A_-^\dagger A_-+1 &  h_{-B}&=-B_+^\dagger B_+ -B_-^\dagger B_- -1 \nonumber\\
h_{A}+&=-(A_+A_- + A_+^\dagger A_-^\dagger) & h_{+B}&=-(B_+B_-+B_+^\dagger B_-^\dagger) \nonumber \\
d_{A}&=i(A_+^\dagger A_-^\dagger -A_+ A_-) &   d_{B}&=i(B_+B_--B_+^\dagger B_-^\dagger)
\end{xalignat}
and the observables  $O_{12}$ and $O_{34}$ as
\ba
O_{12}&=&u_1p_2-p_1u_2=A_+^\dagger A_+ -A_-^\dagger A_- \nonumber \\
O_{34}&=&\pi_1v_2-v_1\pi_2=-B_+^\dagger B_+ +B_-^\dagger B_- \q .
\ea
The (common) eigenspaces (corresponding to the eigenvalues $j,j'\in \Zl$) for these observables are spanned by $\{|k_+,k_-,k'_+,k'_->;k_+-k_-=j \;\text{and}\; k'_--k'_+=j';k_+,k_-,k'_+,k'_-\in \Nl\}$, where $|k_+,k_-,k'_+,k'_->$ denotes a Fock state with respect to the annihilation operators $A_+,A_-,B_+,B_-$. A closer inspection reveals that these eigenspaces are indeed invariant under the $sl(2,\Rl)$-algebra.

The action of the $sl(2,\Rl)$-algebra on each of the above subspaces is a realization of the tensor product representation $D(|j|+1)\otimes D^*(|j'|+1)$ (see \ref{appsl2r}). That can be verified by considering the $h_{-A}$- and the $h_{-B}$-spectrum on these subspaces. The $h_{-A}$-spectrum is bounded from below by $(|j|+1)$, whereas the $h_{-B}$-spectrum is bounded from above by $-(|j'|+1)$. This characterizes $D(|j|+1)$- and $D^*(|j'|+1)$-representations respectively.

Up to now we have achieved
\be
\omega \otimes \omega \otimes \omega^*\otimes\omega^*
= \bigg[D(1)\oplus \sum_{k=2}^{\infty} 2 D(k)\bigg]\otimes\bigg[D^*(1)\oplus \sum_{k=2}^{\infty} 2 D^*(k)\bigg] \q .
\ee 
For a complete reduction of $\omega\otimes\omega\otimes\omega^*\otimes\omega^*$ we have to reduce the tensor products $D(|j|+1)\otimes D^*(|j'|+1)$.

\subsection{The Spectrum of the Master Constraint Operator}

In \cite{neun} the decomposition of all possible tensor products between unitary irreducible representations of $SL(2,R)$ was achieved.

( Actually \cite{neun} considers only representations of $SL(2,\Rl)/\pm \text{Id}$, i.e. representations with uneven $j$ and $j'$. However the results generalize to representations with even $j$ or $j'$. See \cite{repka} for a reduction of all tensor products of $SL(2,\Rl)$, using different methods.)

The strategy in this article is to calculate the spectral decomposition of the Casimir operator. Since the Casimir commutes with $H_-$ one can consider the Casimir operator on each eigenspace of $H_-$. %On these eigenspaces (and in the realization choosen in \ref{neun}) the Casimir acts as an ordinary differential operator. 

Since the Master Constraint Operator is the sum $\MCO=4\Fc+2H_-$ we can easily adapt the results of \cite{neun} for the spectral decomposition of the Master Constraint Operator. In the following we will shortly summarize the results for the spectrum of the Master Constraint Operator. The explicit eigenfunctions are constructed in appendix \ref{appneun}.

To this end we define the subspaces $V(k, j,  j'),\,k\in \Zl,\,|j|\in \Nl$ by
\be
H_-|_{V(k, j,  j')}=k  \q\q \text{and} \q\q
O_{12}|_{V(k, j,  j')}=j \q\q\text{and} \q\q
O_{34}|_{V(k, j,  j')}=j' \,\, .
\ee
$V(k, j,  j')$ is the $H_-$-eigenspace corresponding to the eigenvalue $k$ of the tensor product representation $D(|j|+1) \otimes D^*(|j' |+1)$. Since the $H_-$-spectrum is even for $(j-j')$ even and uneven for $(j-j')$ uneven these subspaces are vacuous for $k+j-j'$ uneven. One result of \cite{neun} is, that the spectrum of the Casimir operator is non-degenerate on these subspaces, which means that there exists a generalized eigenbasis in $\Fl^2(\Rl^4)$ labeled by $(k,j,j')$ and the eigenvalue $\lambda_\Fc$ of the Casimir. 

 The spectrum of the Casimir $\Fc$ on the subspace $V(k, j,  j')$
has a discrete part only if $k>0$ for $|j|-|j'| \geq 2$ or $k<0$ for $|j|-|j'|\leq 2$. There is no discrete part if $||j|-|j'||<2$. The discrete part is for $(j-j')$ and $k$ even
\ba
\lambda_\Fc&=& t(1-t) \quad \text{with} \quad t=1,2,\ldots,\tfrac{1}{2}\text{min}(|k|,||j|-|j'||) \nonumber \\ 
&=& 0,-2,-6,\ldots  \,\,.
\ea
For $(j-j')$ and $k$ odd we have
\ba
\lambda_\Fc&=& t(1-t) \quad \text{with} \quad t=\tfrac{3}{2},\tfrac{5}{2},\ldots,\tfrac{1}{2}\text{min}(|k|,||j|-|j'||) \nonumber \\ 
&=& -\tfrac{3}{4},-\tfrac{15}{4}, -\tfrac{ 35}{4} \ldots  \,\,.
\ea
The continuous part is in all cases the same and given by:
\be
\lambda_\Fc=\tfrac{1}{4}+x^2 \quad \text{with} \quad x \in \big[0,\infty\big) \q .
\ee
%This spectrum is non-degenerate on $V(k, j,  j')$, i.e. there exist only one generalized eigenvector in this space for each generalized eigenvalue of $\Fc$.
The discrete part corresponds to unitary irreducible representations from the positive and negative discrete series of $SL(2,\Rl)$, the continous part corresponds to the (two) principal series of $SL(2,\Rl)$ (see appendix \ref{appsl2r}).

For the spectrum of the Master Constraint Operator we have to multiply with $4$ and add $2k^2$:
\ba
&\lambda_{\MCO}=4 t(1-t)+ 2k^2 \geq 2k^2-k^2+2|k| \quad \nn \\
&\text{with} \quad t=1,2,\ldots,\tfrac{1}{2}\text{min}(|k|,||j|-|j'||)\,\, \text{for even} \,\,k  \nonumber \\
&\text{with} \quad t=\tfrac{3}{2},\tfrac{5}{2} \ldots,\tfrac{1}{2}\text{min}(|k|,||j|-|j'||)\,\, \text{for odd} \,\,k  \nonumber \\
&\text{and}\\
&\lambda_{\MCO}=1+x^2+2k^2>0
\ea
As one can immediately see, the spectrum does not include zero. 
%(This implies, that the trivial representation is not included in the unitary reduction of the tensor products $D(|j|+1)D^*(|j'|+1)$.) 
Since we have no discrete spectrum for $k=0$ the lowest generalized eigenvalue for the master constraint is $1$ from the continuous part.

We will attempt to overcome this problem by introducing a quantum correction to the Master Constraint Operator. Since $1$ is the minimum of the spectrum we substract $1$ \,($\hbar^2$ if units are restored)  from the Master Constraint Operator. 

For the modified Master Constraint Operator we get one solution appearing in the spectral decomposition for each value of $j$ and $j'$. We call this solution $|\lambda_\Fc=\tfrac{1}{4},k=0,j,j'>$ and the linear span of these soltutions $SOL'$.%({\it We assume that these solution are normalized with respect to a scalarproduct, which will be constructed later.}) 
(The above results show, that these quantum numbers are sufficient to label uniquely vectors in the kinematical Hilbert space.)

At the classical level we have several relations between observables, which are valid on the constraint hypersurface. For a physical meaningful quantization we have to check, whether these relations are valid or modified by quantum corrections. 

For our modified Master Constraint Operator this seems not to be the case: At the classical level we have $Q_3=0$ or $P_3=0$. But on $SOL'$, these observables evaluate to:
\ba
Q_3 \, |\lambda_\Fc=\tfrac{1}{4},k=0,j,j'> &=& \tfrac{1}{2} (j-j')\,|\lambda_\Fc=\tfrac{1}{4},k=0,j,j'>  \nonumber \\
P_3 \, |\lambda_\Fc=\tfrac{1}{4},k=0,j,j'> &=& \tfrac{1}{2} (j+j')\,|\lambda_\Fc=\tfrac{1}{4},k=0,j,j'>   \q .
\ea
Since $j$ and $j'$ are arbitrary whole numbers, both $Q_3$ and $P_3$ can have arbitrary large eigenvalues (on the same eigenvector) in $SOL'$.

To solve this problem, we will modify the Master Constraint Operator again, by adding a constraint, which implements the condition $Q_3=0$ or $P_3=0$. Together with the identities (\ref{quid}) this would ensure that $Q_i=0 \, \forall i$ or $P_i=0\, \forall i$ modulo quantum corrections. 

One possibility for the modified constraint is $\MCO''=\MCO-1+(Q_3 P_3)^2$.(This operator is hermitian, since $Q_3$ and $P_3$ commute.) Because of the last relation of (\ref{quid}) this modification can be seen as adding the square (of one quarter) of the right hand side of this relation, i.e. the added part is the square of a linear combination of the constraints.

We already know the spectral resolution of $\MCO''$, since we used $Q_3$ and $P_3$\,\,(or $O_{12}$ and $O_{34}$) in the reduction process for the Master Constraint Operator. Solutions to the Master Constraint Operator $\MCO''$ are the states $|\lambda_\Fc=\tfrac{1}{4},k=0,j,j'>$ with $|j|=|j'|$. We call this solution space $SOL''$. (Up to now this spase is just the linear span of states $|\lambda_\Fc=\tfrac{1}{4},k=0,j,j'>$ with $|j|=|j'|$. Later we will specify a topology for this space.)

Now the observable algebra (\ref{obssl}) does not leave this solution space invariant, since not all observables commute with the added constraint $Q_3 P_3$. 
However, the observables (\ref{obssl}) are redundant on $SOL''$, since they obey the relations (\ref{quid}). So the question is, whether one can find enough observables, which commute with $Q_3 P_3$ (and with the constraints, we started with) to carry all relevant physical information. 
Apart from $Q_3$ and $P_3$ the operators $Q_1^2+Q_2^2$ and $P_1^2+P_2^2$ commute with the added constraint. But the latter do not carry additional information about physical states, because of the first relation in (\ref{quid}). Operators of the form $p_1(Q)Q_3 +p_2(P) P_3$, where $p_1(Q)$ (resp. $p_2(P)$) represents a polynomial in the $Q$-observables ($P$-observables), commute with $Q_3 P_3$ on the subspace defined by $Q_3 P_3=0$.  Likewise operators of the form $p_1(Q)|\text{sgn}(Q_3)|+p_2(P)|\text{sgn}(P_3)|$, where $\text{sgn}$ has values $1,0$ and $-1$ (and is defined by the spectral theorem) leave $SOL''$ (formally) invariant.  

 In the following we will take as observable algebra the algebra generated 
 by the elementary operators $|\text{sgn}(Q_3)|Q_i|\text{sgn}(Q_3)|$ and 
 $|\text{sgn}(P_3)|P_i|\text{sgn}(P_3)|$. This algebra is closed under 
 taking adjoints. Notice, however, that we may add operators such
as $|\text{sgn}(Q_3)|Q_+ Q_+|\text{sgn}(Q_3)|$ which does not leave the 
sectors invariant and thus destroy the superselection structure which is 
a physical difference from the results of \cite{louko}. 
The next section shows that the latter operator transforms states from the sector 
$\{\text{sgn}(Q_3)=-1,\text{sgn}(P_3)=0\}$ 
to the sector $\{\text{sgn}(Q_3)=+1,\text{sgn}(P_3)=0\}$ 
(since $Q_\pm,P_\pm$ are ladder operators, 
which raise or lower the $Q_3,P_3$ eigenvalues by $1$ respectively). 
Likewise one can construct operators wich transform from the sector 
$\{\text{sgn}(P_3)=0\}$ to $\{\text{sgn}(Q_3)=0\}$ 
and vice versa: For instance 
\be
|\text{sgn}(P_3)|\,P_+ \cdots P_+  \,(1-|\text{sgn}(Q_3)|)(1-|\text{sgn}(P_3)|)\,Q_+\cdots Q_+\,|\text{sgn}(Q_3)|
\ee   
has this property and leaves the solution space to the modified Master 
Constraint Operator invariant. Its adjoint is of the same form, 
transforming from the sector $\{\text{sgn}(Q_3)=0\}$ to the 
sector $\{\text{sgn}(P_3)=0\}$. Thus we may map between all five sectors 
mentioned before except for the origin. There seems to be 
no natural exclusion principle for these operators from the point of view
of DID and thus we should take them seriously.

\subsection{The Physical Hilbert space}
\label{physhilbert}
Now one can use the spectral measure for the Master Constraint Operator and construct a scalar product in $SOL'$ and $SOL''$ and then complete them into Hilbert spaces $\Fh'$ and $\Fh''$. This is done explicitly in Appendix \ref{appneun}, here we only need that this can be done in principle.

The so achieved Hilbert space $\Fh'$ has to carry a unitary representation of the observable algebra $sl(2,\Rl)\times sl(2,\Rl)$ (since these observables commute with the constraints). In particular we already know the spectra of $Q_3$ and $P_3$ to be the integers $\Zl$, since we diagonalized them simultaneously with the Master Constraint Operator. These spectra are discrete, which means that in the constructed scalar product the states $|\lambda_\Fc=\tfrac{1}{4},k=0,j,j'>$ (which are eigenstates for $Q_3$ and $P_3$) have a finite norm. So we can normalize them to states $||j,j'>>$ and in this way obtain a basis of $\Fh'$.    

Now, because of the identity (\ref{quid}) we also know the value of the $sl(2,\Rl)$ Casimirs $Q_1^2+Q_2^2-Q_3^2$ and $P_1^2+P_2^2-P_3^2$ on $\Fh'$ to be $\tfrac{1}{4}$. Together with the fact that $\Fh'$ has a normalized eigenbasis  $\{||j,j'>>,j,j'\in \Zl\}$ (with respect to $Q_3$ and $P_3$) we can determine the unitary representation of $sl(2,\Rl)\times sl(2,\Rl)$ to be $P(t=1/2,\eps=0)_Q \otimes P(t=1/2,\eps=0)_P$ (see Appendix \ref{appsl2r}). This fixes the action of the (primary) observable algebra to be (modulo phase factors, which can be made to unity by adjusting phases of the basis vectors):
\begin{alignat}{2}
&Q_+ ||j,j'>> \q&=&\q\tfrac{1}{2\sqrt{2}}((j-j')+ 1)\,||j + 1,j'-1>>  \nonumber \\
&Q_-||j,j'>> \q&=& \q\tfrac{1}{2\sqrt{2}}((j-j') - 1)\,||j - 1,j'+1>> \nn \\
&Q_3||j,j'>>\q&=&\q\tfrac{1}{2}(j-j')\, ||j,j'>> \nonumber \\
&P_+ ||j,j'>>\q&=&\q\frac{1}{2\sqrt{2}}((j+j') + 1)\,||j+1,j' + 1>>\nonumber \\ 
&P_- ||j,j'>>\q&=&\q\frac{1}{2\sqrt{2}}((j+j') - 1)\,||j-1,j'-1>> \nonumber \\
&P_3||j,j'>>\q&=&\q\tfrac{1}{2}(j+j')|\,|j,j'>> 
\end{alignat}
From these results we can derive the action of the altered observable algebra on $\Fh''$, i.e. on states $||j,\eps j>>$ with $\eps=\pm 1$:
\begin{alignat}{2}  \label{alterobs}
&\Theta (Q_3) Q_+ \Theta (Q_3)\,||j,\eps j>>\q &=& \q \delta_{-1,\eps}\,(1-\delta_{-1,j})\,\tfrac{1}{\sqrt{2}}(j+ \tfrac{1}{2})\,||j + 1,\eps (j+1)>>  \nonumber \\
&\Theta (Q_3) Q_-\Theta (Q_3) \,||j,j'>> \q &=& \q \delta_{-1,\eps}\,(1-\delta_{+1,j})\,\tfrac{1}{\sqrt{2}}(j- \tfrac{1}{2})\,||j - 1,\eps(j-1)>> \nn \\
&Q_3\,||j,\eps j>>  \q &=& \q \delta_{-1,\eps}\, j\,||j,\eps j>> \nonumber \\
&\Theta (P_3) P_+\Theta (P_3) \,||j,j'>>  \q &=& \q \delta_{1,\eps}\,(1-\delta_{-1,j})\,\tfrac{1}{\sqrt{2}}\,(j+ \tfrac{1}{2})\, ||j+1,\eps(j + 1)>>\nonumber \\ 
&\Theta (P_3) P_-\Theta (P_3) \, ||j,j'>> \q &=& \q \delta_{1,\eps}\,(1-\delta_{+1,j})\,\tfrac{1}{\sqrt{2}}(j- \tfrac{1}{2})\, ||j-1,\eps(j-1)>> \nonumber \\
&P_3\, ||j,\eps j>> \q      & =& \q \delta_{1,\eps}\,j\,||j,\eps j>> 
\end{alignat}
where we abbreviated $|\text{sgn}(\bold O)|$ by $\Theta(\bold O)$.
The state $|0,0>$ is annihilated by all (altered) observables.

\subsection{Algebraic Quantization}

 In \cite{louko} the $SL(2,\Rl)$-model has been quantized in the Algebraic 
 and Refined Algebraic Quantization framework. We will shortly review the 
 results of the Algebraic Quantization scheme in order to compare them 
with the Master Constaint Programme.

 In this scheme one starts with the auxilary Hilbert space $L^2(\Rl^4)$, 
 the constraints (\ref{qsl}) and a $^*$- algebra of observables $\ca^*$. 
 One looks for a solution space for the constraints which carries an 
irreducible representation of $\ca^*$ and for a scalar product on this space in which the star-operation becomes the adjoint operation.

The solution space $\tilde V$, which was found in \cite{louko} is the linear span of states $|j,\eps j>$ where $j$ is in $\Zl$ and $\eps \in \{-1,+1\}$. These states are expressible as smooth functions on the $(\vec{u},\vec{v})$ configuration space $\Rl^4$ and they solve the constraints (\ref{qsl}). 

 The solution states can be expressed in our ``polarized'' Fock basis as follows  
\ba
|j,\eps j>=\sum_{m=0}(-1)^m |m+\tfrac{1}{2}(j+|j|),m+\tfrac{1}{2}(-j+|j|)>\otimes \nn \\|m+\tfrac{1}{2}(-\eps j+|j|),m+\tfrac{1}{2}(\eps j+|j|)> \,\,.
\ea
(These states are the solutions $f(t=1;k=0,j,j'=\eps j)$, see (\ref{appneun}).)
Clearly, the states $|j, \eps j>$ solve the Master Constraint Operator $\MCO$. However there are much more solutions to the Master Constraint Operator (which do not necessarily solve the three constraints (\ref{qsl})).

The algebra $\ca^*$ used in \cite{louko} is the algebra generated by the observables (\ref{obssl}). The star-operation is defined by $Q_i^*=Q_i,\,\,P_i^*=P_i$ and extended to the full algebra by complex anti-linearity. This algebra is supplemented to the algebra $\ca^*_{ext}$ by the operators $R_{\eps_1,\eps_2}=R^*_{\eps_1,\eps_2}$, which permute between the four different sectors of the classical constraint phase space:
\be 
R_{\eps_1,\eps_2}:(u_1,u_2,v_1,v_2,p_1,p_2,\pi_1,\pi_2)\mapsto (u_1,\eps_1 u_2,v_1,\eps_1 \eps_2 v_2,p_1,\eps_1 p_2,\pi_1,\eps_1 \eps_2 \pi_2)
\ee

The algebra $\ca^*$ has the following representation on $\tilde V$: 
\ba \label{repb}
Q_3|j,\eps j>&=&\delta_{-1,\eps}\, j\, |j,\eps j> \nn \\
Q_\pm |j,\eps j>&=& \delta_{-1,\eps} \,(\tfrac{\mp i}{\sqrt{2}}|j|)\,|(j\pm 1),\eps\,(j\pm 1)>  \nn \\
P_3|j,\eps j>&=&\delta_{+1,\eps} \,j\, |j,\eps j> \nn \\
P_\pm |j,\eps j>&=& \delta_{+1,\eps} \,(\tfrac{\mp i}{\sqrt{2}}|j| ) \,|(j\pm 1), \eps\,(j\pm 1)>  
\ea
The state $|j=0,j'=0>$ is annihilated by all operators in $\ca^*$, in particular, it generates an invariant subspace for $\ca^*$. Now, it is not possible to introduce an inner product on $\tilde V$, in which the star-operation becomes the adjoint operation (because the $SO(2,2)$-representation defined by (\ref{repb}) is non-unitary). However, since $|0,0>$ generates an invariant subspace one can take the quotient $\tilde{V}/\{c|0,0>,c\in \Cl\}$, consisting of equivalence classes $\big[ v\big] =\{v + c\,|0,0>,\,\, c\in \Cl\}$ where $v \in \tilde{V}$. (In particular $\big[|0,0>\big]$ is the null vector $\big[0\big]$.) The $so(2,2)$-representation on $\tilde{V}$ then defines a representation on this quotient space by $O([v])=[O(v)]$, where $O$ is an $so(2,2)$-operator. (This representation is well defined because we are quotienting out an invariant subspace.) A basis in this quotient space is $\{\big[|j,\eps j>\big],j\in \Nl-\{0\}\}$. In the following we will drop the equivalence class brackets $[\cdot]$.

 The quotient representation is the direct sum of four (unitary) irreducible representations of $so(2,2)$, labeled by $\text{sgn}(j)=\pm 1$ and $\eps=\pm 1$. The inner product, which makes these representations unitary is
\be
<j_1,\eps_1 j_1 |j_2,\eps_2 j_2>=c(\text{sgn}(j),\eps)\,\delta_{j_1,j_2}\,\delta_{\eps_1,\eps_2}\,|j|  
\ee
where $c(\text{sgn}(j),\eps)$ are four independent positiv constants.

  By taking the reflections $R_{\eps_1,\eps_2}\in O(2,2)$ into account, we can partially fix these constants. Their action on states in $\Fl^2(\Rl^4)$ is
\be
 (R_{\eps_1,\eps_2}\psi)(u_1,u_2,v_1,v_2)=\psi(u_1,\eps_1 u_2,v_1,\eps_1 \eps_2 v_2) \q .
\ee
States with angular momenta $j$ and $j'$ are mapped to states with angular momenta $\eps_1 j$ and $\eps_1\eps_2 j'$. Therefore the $R_{\eps_1,\eps_2}$'s effect, that the quotient representation of the observable algebra becomes an irreducible one. Since $R_{\eps_1,\eps_2}$ is in $O(2,2)$, it is a natural requirement for them to act by unitary operators. This fixes the four constants $c(\text{sgn}(j),\eps)$ to be equal (and in the following we will set them to $1$).

This gives for the action of the algebra $\ca^*$ on the normalized basis vectors $|j, \eps j>_N:=\frac{1}{\sqrt{|j|}}|j,\eps\,j>,\,\, j \in \Zl-\{0\}$:
\ba \label{algquobs}
Q_3|j,\eps j>_N &=&\delta_{-1,\eps}\, j\,\, |j,\eps j>_N \nn \\
Q_\pm |j,\eps j>_N &=& \delta_{-1,\eps} \,(\tfrac{\mp i}{\sqrt{2}}\sqrt{|j(j\pm 1)|})\,\,|(j\pm 1),\eps\,(j\pm 1)>_N  \nn \\
P_3|j,\eps j>_N &=&\delta_{+1,\eps} \,j\,\, |j,\eps j>_N \nn \\
P_\pm |j,\eps j>_N &=& \delta_{+1,\eps} \,(\tfrac{\mp i}{\sqrt{2}}\sqrt{|j(j\pm 1)|} ) \,\, |(j\pm 1), \eps\,(j\pm 1)>_N  \q.
\ea
In the limit of large $j$ the right hand sides of (\ref{alterobs}) and (\ref{algquobs}), ie. the matrix elements of the observables in the two qunatizations, coincide except for phase factors. These can be made equal by adjusting the phase factors of the respective basic vectors. Therefore both quantization programs lead to the same semiclassical limit.

 A first crucial difference in the results of the two quantization 
approaches is 
 that in the Master Constraint Programme the vector $||j=0,j'=0>>$ is 
 included in the physical Hilbert space whereas it is excluded during the 
 Algebraic Quantization process. If we exclude the sector changing 
operators mentioned above by hand, then $||j=0,j'=0>>$ is annihilated by 
 the altered observable algebra and likewise cannot be reached by applying 
observables to other states in the physical Hilbert space. If we include 
the sector changing operators then $|j=0,j'=0>$ is still not in the range 
of any observable because the observables are sandwiched beween operators 
of the form $|\mbox{sgn}(Q_3)|,\;|\mbox{sgn}(Q_3)|$. However, one can map 
between all the remaining sectors which thus provides a second difference 
with \cite{louko}.

%normal ordering of the master constraint
% :M:=M-3A_+^\dagger A_+ -3A_-^\dagger A_--3B_+^\dagger B_+ -3B_-^\dagger B_--4

\section{Conclusions}
\label{s9}

What we learnt in this paper is that the Master Constraint Programme can 
also successfully be applied to the difficult of constraint algebras 
generating non -- amenable, non -- compact gauge groups. As was observed 
for instance in \cite{gomberoff} this is a complication which affects 
the group averaging proposal \cite {1.10}
for solving the quantum constraints
quite drastically in the sense that the physical Hilbert space depends
critically on the choice of a dense subspace of the Hilbert space. 
The Master Constraint Programme also faces complications: The spectrum
is supported on a genuine subset of the positive real line not containing 
zero. Our proposal to subract the zero point of the spectrum from the 
Master Constraint, which can be considered as a quantum 
correction\footnote{The fact that the correction is quadratic in $\hbar$ 
rather than linear in contrast to the normal ordering correction of the 
harmonic osciallor can be traced back to the fact that harmonic oscillator
Hamiltonian can be considered the Master Constraint for the second class 
pair of constraints $p=q=0$ while the $sl(2R)$ constraints are first 
class.} because it is proportional 
to $\hbar^2$ worked and produced an acceptable physical Hilbert space.

Of course, it is unclear whether that physical Hilbert space is in a sense 
the only correct choice because the models discussed are themselves not 
very physical and therefore we have only mathematical consistency as a 
selection criterion at our disposal, such as the fact that the algebraic
approach reaches the same semiclassical limit by an independent method.
Nevertheless, it is important to 
notice that DID produces somewhat different results than algebraic and
RAQ methods, in particular, the superselection theory is typically 
trivial in contrast to those programmes. It would be good to
know the deeper or intuitive reason behind this and other differences. 
Obviously,
further work on non -- amenable groups is necessary, preferrably in an 
example which has a physical interpretation, in order to 
settle these interesting questions.\\
\\
\\
\\
\\
{\large Acknowledgements}\\
\\
We thank Hans Kastrup for fruitful discussions about $SL(2,\Rl)$, 
especially for pointing out reference \cite{neun}. 
BD thanks the German National Merit Foundation for financial support.
This research project was supported in part by a grant from
NSERC of Canada to the Perimeter Institute for Theoretical Physics.

\begin{appendix}

\section{Review of the Representation Theory of $SL(2,\Rl)$ and its 
various Covering Groups}
\label{sa}

\subsection{ $sl(2,\Rl)$ Representations}
\label{appsl2r}

In this section we will review unitary representations of $sl(2,\Rl)$, see \cite{bargmann,adams}. 

In the defining two-dimensional representation the $sl(2,\Rl)$-algebra is spanned by
\ba
h= \frac{-1}{2i} \left( \begin{array}{cc}
                                        0 & 1     \\
                                        -1 & 0   
                                        \end {array} \right)\q\q
 n_1= \frac{-1}{2i} \left( \begin{array}{cc}
                                        0 & 1     \\
                                        1 & 0   
                                        \end {array} \right)\q\q
n_2= \frac{-1}{2i} \left( \begin{array}{cc}
                                        1 & 0     \\
                                        0 & -1   
                                        \end {array} \right)
\ea
with commutation relations
\ba
\big[h,n_1\big]=in_2 \q \q \big[n_2,h\big]=in_1 \q \q \big[n_1,n_2\big]=-ih \q.
\ea
We introduce raising and lowering operators $n^{\pm}$ as complex linear combinations $ n^{\pm}=n_1 \pm i n_2$ of  $n_1$ and $n_2$, which fulfill the algebra 
\ba \label{commis1}
\big[h,n^\pm \big]=\pm n^\pm \q \q  \big[n^+,n^-\big]=-2h \q .
\ea
The Casimir operator, which commutes with all $sl(2,\Rl)$-algebra operators, is
\be \label{casi1}
\Fc=-h^2+\tfrac{1}{2}(n^+ n^-+n^-n^+)=-h^2+n_1^2+n_2^2 \q.
\ee

We are interested in unitary irreducible representations of $sl(2,\Rl)$, i.e. representations where $h,n_1$ and $n_2$ act by self-adjoint operators on a Hilbert space, which does not have non-trivial subspaces, that are left invariant by the $sl(2,\Rl)$-operators. According to Schur's Lemma the Casimir operator acts on an irreducible space as a multiple of the identity operator $\Fc=\fc\,\text{Id}$. Since $n_1$ and $n_2$ are self-adjoint operators, the raising operator $n^+$ is the adjoint of the lowering operator $n^-$ and vice versa. (For notational convenience we often do not discriminate between elements of the algebra and the operators representing them.) 

In general the $sl(2,\Rl)$-representations do not exponentiate to a representation of the group $SL(2,\Rl)$ but to the universal covering group $\tilde{SL(2,\Rl)}$. Since $h$ is the generator of the compact subgroup of $\tilde{SL(2,\Rl)}$ it will have discrete spectrum (and therefore normalizable eigenvectors) in a unitary representation. If the $sl(2,\Rl)$-representation exponentiates to an $SL(2,\Rl)$ representation, $h$ has spectrum in $\{\tfrac{1}{2}n,n\in \Zl\}$. If in this group representation the center ${\pm \text{Id}}$ acts trivially, it is also an $SO(2,1)$ representation, since $SO(2,1)$ is isomorphic to the quotient group $SL(2,\Rl)/\{\pm \text{Id}\}$. In this case $h$ has spectrum in $\Zl$. 

Now, assume that $|\fh>$ is an eigenvector of $h$ with eigenvalue $\fh$. Using the commutation relations (\ref{commis1}) one can see, that $n^\pm |\fh>$ is either zero or an eigenvector of $h$ with eigenvalue $\fh \pm 1$. By repeated application of $n^+$ or $n^-$ to $|\fh>$ one therefore obtains a set of eigenvectors $\{|\fh+n>\}$ and corresponding eigenvalues $\{\fh+n\}$, where $n$ is an integer. This set of eigenvalues may or may not be bounded from above or below. 

Similarly, one can deduce from the commutation relations that $n^+ n^- |\fh>$ and $n^- n^+ |\fh>$ are both eigenvectors of $h$ with eigenvalue $\fh$ (or zero). Apriori these eigenvectors do not have to be a multiple of $|\fh>$, since it may be, that $h$ has degenerate spectrum. But this is excluded by the relations
\be \label{commi2}
n^+n^-=h^2-h+\Fc \q\q\q  n^-n^+=h^2+h+\Fc \q,
\ee
obtained by using (\ref{commis1},\ref{casi1}). (Remember, that $h$ and $\Fc$ act as multiples of the identity on $|\fh>$.) From this one can conclude that the set $\{|\fh+n>\}$ is invariant under the $sl(2,\Rl)$-algebra (modulo multiples) and hence can be taken as a complete basis of the representation space.

One can use the relations (\ref{commi2}) to set constraints on possible eigenvalues of $h$ and $\Fc$. Consider the scalar products
\ba
<\fh \pm 1 |\fh \pm 1>&=&  <\fh|(n^\pm)^\dagger n^\pm |\fh>=<\fh|n^\mp n^\pm |\fh>   \nn \\ 
&=&<\fh| h^2 \pm h+\Fc   |\fh> =(\fh^2 \pm \fh +\fc)<\fh|\fh> \; .\q\q 
\ea 
Since the norm of a vector has to be positive one obtains the inequalities
\be \label{ineq}
\fh^2 \pm \fh +\fc  \geq 0
\ee
for the spectrum of $h$ and the value of the Casimir $\Fc=\fc\,\text{Id}$. 

To summarize what we have said so far, we can specify a unitary irreducible representation with the help of the spectrum of $h$ and the eigenvalue of the Casimir $\fc$. The spectrum of $h$ is non-degenerate and may be unbounded or bounded from below or from above. Together with $\fc$ the spectrum has to fulfill the inequalities (\ref{ineq}). In this way one can find the following irreducible representations of $sl(2,\Rl)$ (For an explicit description, how one can find the allowed representation parameters, see \cite{bargmann,adams}):
\begin{itemize}
\item[(a)] The principal series $P(t,\eps)$ where $t\in\{\tfrac{1}{2}+ix,x\in\Rl\wedge x\geq 0\}$ and $\eps=\fh(\text{mod}\,1)\in (-\tfrac{1}{2},\tfrac{1}{2}\big]$. 

The spectrum of $h$ is unbounded and given by $\{\eps +n,n\in \Zl\}$. The Casimir eigenvalue is $\fc=t\,(1-t)\geq\tfrac{1}{4}$. (For $t=\tfrac{1}{2},\eps=\tfrac{1}{2}$ the representation $P(t,\eps)$ is reducible into $D(1)$ and $D^*(1)$ see below.) 

\item[(b)] The complementary series $P_c(t,\eps)$ where $\tfrac{1}{2}< t<1$ and $|\eps|<1-t$.   

The spectrum of $h$ is unbounded and given by $\{\eps +n,n\in \Zl\}$. The Casimir eigenvalue is $0<\fc=t\,(1-t)<\tfrac{1}{4}$

\item[(c)] The positive discrete series $D(k)$ where $k>0$.

Here, the spectrum of $h$ is bounded from below by $\tfrac{1}{2}k$ and we have $\text{spec}(h)=\{\tfrac{1}{2}k+n,n\in \Nl\}$. The value of the Casimir is $\fc=\tfrac{1}{2} k- \tfrac{1}{4}k^2 \leq \tfrac{1}{4}$.

\item[(d)] The negative discrete series $D^*(k)$ where $k>0$.

In this case the spectrum of $h$ is bounded from above by $-\tfrac{1}{2}k$ and we have $\text{spec}(h)=\{-\tfrac{1}{2}k-n,n\in \Nl\}$. The value of the Casimir is $\fc=\tfrac{1}{2} k-\tfrac{1}{4} k^2\leq \tfrac{1}{4}$.

\item[(e)] The trivial representation.

\item[] As mentioned above representations with an integral $h$-spectrum can be exponentiated to representations of the group $SO(2,1)$, if the spectrum includes half integers one obtains representations of $SL(2,\Rl)$ and for $\text{spec}(h)\in \{\frac{1}{4}n,n\in \Zl\}$ representations of the double cover of $SL(2,\Rl)$ (the metaplectic group).  

\end{itemize}    

Finally we want to show, how one can uniquely determine the action of the $sl(2,\Rl)$-algebra in a representation from the principal series. (The other cases are analogous, but we need this case in section \ref{physhilbert}.) To this end we assume that the vectors $||\fh>>$ are normalized eigenvectors of $h$ with eigenvalue $\fh$. Applying $n^\pm$ gives a multiple of $||\fh \pm 1>>$:
\be
n^\pm||\fh>> =A_\pm(\fh)||\fh\pm 1>> \q.
\ee

Using relation (\ref{commi2}) one obtains for the coefficients $A_\pm(\fh)$
\ba
n^\mp n^\pm ||\fh>>=A_\mp(\fh\pm1)A_\pm(\fh)|\fh>>=( \fh^2\pm\fh+\fc)||\fh>>\q . 
\ea
Furthermore
\be
A_+(\fh)=<<\fh+1||n^+||\fh>>=\overline{<<\fh||n^-||\fh+1>>}=\overline{A_-(\fh+1)} \q.
\ee
The solution to these equations is 
\ba
A_+=c_+(\fh)\,(\fh+t) \q \q A_-(\fh)=c_-(\fh)\,(\fh-t)
\ea
where $|c_\pm|=1$ and $c_+(\fh)c_-(\fh+1)=1$. Solutions with different $c_\pm$ are related by a phase change for the states $||\fh>>$.

\subsection{ Oscillator Representations}
\label{apposci}
\subsubsection{The Oscillator Representation}

Here we will summarize some facts about oscillator representations, following \cite{howe,howeart}.

The oscillator representation is a unitary representation of $\widetilde{SL(2,\Rl)}$ the double cover of $SL(2,\Rl)$ (the so-called metaplectic group) and is also known under the names Weil representation, Segal-Shale-Weil representation or harmonic representation. 

The associated representation $\omega$ of the Lie algebra $sl(2,\Rl)$ on $\Fl^2(\Rl)$ is given by
\begin{alignat}{2} \label{osci1}
h=\tfrac{1}{2}(a^\dagger a +\tfrac{1}{2}) \q\q
&n_1=\tfrac{1}{4}(a^\dagger a^\dagger+a a) \q \q && n_2=-i\tfrac{1}{4}(a^\dagger a^\dagger-a a) \nn \\
&n^+=n_1+in_2=\tfrac{1}{2}a^\dagger a^\dagger  \q \q && n^-=n_1-in_2=\tfrac{1}{2}a a  % \nn \\
%\big[h,n_1\big]=in_2 \q \q \big[n_2,h\big]=in_1 \q \q \big[n_1,n_2\big]=-ih \nn \\
%\big[h,n^\pm \big]=\pm n^\pm \q \q  \big[n^+,n^-\big]=-2h \q \q n^+^\dagger=n^-
\end{alignat}
where we introduced annihilation and creation operators
\be
a=\tfrac{1}{\sqrt{2}}(x+ip) \q\text{and} \q a^\dagger=\tfrac{1}{\sqrt{2}}(x-ip)\q \text{with} \q p=-i\tfrac{d}{dx} \q .
\ee  

The operator $h$ is (half of) the harmonic oscillator Hamiltonian and represents the infinitesimal generator of the two-fold covering group of $SO(2)$. It has discrete spectrum $\text{spec}(h)= \{\tfrac{1}{4}+\tfrac{1}{2}n,n\in \Nl\}$ and its eigenstates are the Fock states $|n>=(n!)^{-1/2}(a^\dagger)^n|0>;\,a|0>=0$, which form an orthonormal basis of $\Fl^2(\Rl)$.
As can be easily seen, the representation (\ref{osci1}) leaves the spaces 
of even and odd number Fock states invariant, therefore the 
representation 
is reducible into two subspaces. 
These subspaces are irreducibel since one can reach each  (un-)even Fock 
state $|n>$  by applying powers of $n^+$ or $n^-$ to an arbitrary 
(un-)even Fock state $|n'>$. 
 
Since we have an $h$-spectrum which is bounded from below by $\tfrac{1}{4}$ for the even number states and $\tfrac{3}{4}$ for the uneven number states, the corresponding representations are $D(1/2)$ and $D(3/2)$ from the positive discrete series (of the metaplectic group). 
  
The Casimir of the oscillator representation is a constant:
\be 
\Fc(\omega)=-h^2+\tfrac{1}{2}(n^+ n^-+n^-n^+)=\tfrac{3}{16} \q.
\ee
This confirms the finding $\omega \simeq D(1/2) \oplus D(3/2)$, since we have 
\be
\Fc(D(1/2))=\tfrac{1}{4}\tfrac{1}{2}(2-\tfrac{1}{2})=\tfrac{3}{16}=\tfrac{1}{4}\tfrac{3}{2}(2-\tfrac{3}{2})=\Fc(D(3/2)) \q.
\ee

In chapter \ref{s5.3} we use an $sl(2,\Rl)$-basis $\{h_{-u_i}=2h,\,h_{+u_i}=2n_1,\,d=2n_2\}$. If one would rewrite the $(u_i)$-representation (\ref{uivi1}) in terms of $\{h,n_1,n_2\}$ it would differ from the representation (\ref{osci1}) by minus signs in $n_1$ and $n_2$. Nevertheless the $(u_i)$-representation is unitarily equivalent to the representation (\ref{osci1}), where the unitary map is given by the Fourier transform $\Ff$. This can be easily seen by using the transformation properties of annihilation and creation operators under Fourier transformation:\,$\Ff a\Ff^{-1}=ia $ and $\Ff a^\dagger\Ff^{-1} =-i a^\dagger$.
%\Ff|n>=(-i)^n |n>    

\subsubsection{Contragredient Oscillator Representations}

The contragredient oscillator representation $\omega^*$ on $\Fl^2(\Rl)$ is given by
\begin{alignat}{2}
h^*=-\tfrac{1}{2}(a^\dagger a +\tfrac{1}{2}) \q\q
 n_1^*&=-\tfrac{1}{4}(a^\dagger a^\dagger+a a) \q \q & n_2^*&=-i\tfrac{1}{4}(a^\dagger a^\dagger-a a) \nn \\
 n^{+*}&=n_1^*+in_2^*=-\tfrac{1}{2}a a  \q \q & n^{-*}&=n_1^*-in_2^*=-\tfrac{1}{2}a^\dagger a^\dagger   
\end{alignat}
Here, $h^*$ has strictly negativ spectrum $\text{spec}(h^*)=  \{-\tfrac{1}{4}-\tfrac{1}{2}n,n\in \Nl\}$.
An analogous discussion to the one above reveals that the contragredient oscillator representation is the direct sum of the representations $D^*(1/2)$ and $D^*(3/2)$ from the negative discrete series (of the metaplectic group). The Casimir evaluates to the same constant as above $\Fc(\omega^*)=\tfrac{3}{16}=\Fc(D^*(1/2)=\Fc(D^*(3/2))$. 

\subsubsection{Tensor Products of Oscillator Representations}
\label{apposcitensor}
Now one can consider the tensor product of $p$ oscillator and $q$ contragredient oscillator representations. The tensor product $(\otimes_{\!p}\omega) \otimes (\otimes_{\!q}\omega^*)$ will be abbreviated by $\omega^{(p,q)}$. The representation space is $(\otimes_{\!p}\Fl^2(\Rl))\otimes (\otimes_{\!q}\Fl^2(\Rl))$ wich can be identified with $\Fl^2(\Rl^{p+q})$. The tensor product representation (of the $sl(2,\Rl)$-algebra) is given by 
\ba \label{hn}
h^{(p,q)}&=&\tfrac{1}{2}\sum_{j=1}^{p}(a_j^\dagger a_j +\tfrac{1}{2})    -  \tfrac{1}{2}\sum_{j=p+1}^{p+q}(a_j^\dagger a_j +\tfrac{1}{2})    \nn \\
n_1^{(p,q)}&=&\tfrac{1}{4} \sum_{j=1}^{p}(a_j^\dagger a_j^\dagger+a_j a_j)   -\tfrac{1}{4}  \sum_{j=p+1}^{p+q}(a_j^\dagger a_j^\dagger+a_j a_j) \nn \\
n_2^{(p,q)}&=& -\tfrac{i}{4}  \sum_{j=1}^{p+q}(a_j^\dagger a_j^\dagger-a_j a_j) \ea    
where $a_j$ and $a_j^\dagger$ denote annihilation and creation operators for the $j$-th coordinate in $\Rl^{p+q}$:
\be
a_j=\tfrac{1}{\sqrt{2}}(x^j+ip_j) \q\text{and} \q a^\dagger=\tfrac{1}{\sqrt{2}}(x^j-ip_j)\q \text{with} \q p_j=-i\partial_j \q .
\ee

On $\Fl^2(\Rl^{p+q})$ we also have a natural action of the generalized orthogonal group $O(p,q)$ given by $g\cdot f(\vec x)=f(g^{-1}(\vec x))$, where $g \in O(p,q)$ and $\vec x \in \Rl^{p+q}$. This action commutes with the $sl(2,\Rl)$ action, which can rapidly be seen, if we calculate the following $sl(2,\Rl)$ basis:   
\ba \label{ed}
e^+&=&2h^{(p,q)}+2n_1^{(p,q)}=\sum_{j=1}^p (x^j)^2-\sum_{j=p+1}^{p+q} (x^j)^2=g_{ij} x^i x^j \nn \\
e^-&=&2h^{(p,q)}-2n_1^{(p,q)} =  \sum_{j=1}^p   (p_j)^2-\sum_{j=p+1}^{p+q} (p_j)^2=g^{ij}p_i p_j   \nn \\
d &=& -2n_2^{(p,q)}  =\tfrac{1}{2}\sum_{j=1}^{p+q} (x^jp_j+p_j x^j)=\tfrac{1}{2}g^i_j(x^jp_i+p_ix^j) \q.
\ea
Here $g^{ij}$ is inverse to the metric $g_{ij}=\text{diag}(+1,\ldots,+1,-1,\ldots,-1)$ (with $p$ positive and $q$ negative entries) so that $g^i_j:=g^{ik}g_{kj}=\delta^i_j$, where in the last formula and in the right hand sides of (\ref{ed})  we summed over repeated indices. Since $O(p,q)$ leaves by definition the metric $g_{ij}$ invariant, the $sl(2,\Rl)$-operators (\ref{ed}) (and all their linear combinations) are left invariant by the $O(p,q)$-action: $\rho(g^{-1})s\hat \rho(g) =s$, where $s$ is an element from the $sl(2,\Rl)$-algebra representation and $\rho( g)$ denotes the action of $g\in O(p,q)$ on states in $\Fl^2(\Rl^{p+q})$.

The action of $O(p,q)$ induces a unitary representation $\rho$ of $O(p,q)$ on $\Fl^2(\Rl^{p+q})$ (defined by $\rho(g)f(\vec x)=f(g^{-1}(\vec x))$ for $f(\vec x)\in\Fl^2(\Rl^{p+q})$). The derived representation of the Lie algebra $so(p,q)$ is given by:
\ba
A_{jk}&=&x^j p_k-x^k p_j, \q j,k=1,\ldots ,p \\
B_{jk}&=&x^j p_k-x^k p_j, \q j,k=p+1,\ldots ,p+q \\
C_{jk}&=&x^j p_k+x^k p_j, \q j=1,\ldots p,\,\,k=p+1,\ldots ,p+q \q.
\ea
The operators $A_{jk}$ and $B_{jk}$ span the Lie algebra $so(p)\times so(q)$ of the maximal compact group $O(p)\times O(q)$. (From this one can conclude that $A_{jk}$ and $B_{jk}$ have discrete spectra.)

For the representations $\rho(O(p,q)$ and $\omega^{(p,q)}(\widetilde{SL(2,\Rl)})$ there is a remarkable theorem, which we will cite from \cite{howeart}:

{\it 
The groups of operators $\rho(O(p,q))$ and $\omega^{(p,q)}(\tilde{SL(2,\Rl)})$ generate each other commutants in the sense of von Neumann algebras. Thus there is a direct integral decomposition
\be \label{hdecomp}
\Fl^2(\Rl^{p+q}) \simeq \int \sigma_s \otimes \tau_s\, ds
\ee
where $ds$ is a Borel measure on the unitary dual of $\widetilde{SL(2,\Rl)}$, and $\sigma_s$ and $\tau_s$ are irreducible representations of $O(p,q)$ and $\widetilde{SL(2,\Rl)}$, respectively. Moreover $\sigma_s$ and $\tau_s$ determine each other almost everywhere with respect to $ds$.
}

This means, that if we are interested in the decomposition of the $\rho(O(p,q))$-representation, we can equally well decompose $\widetilde{SL(2,\Rl)}$, which may be an easier task. (We used this in example \ref{s5.2}.) 

Furthermore, this theorem is very helpful if one of the two group algebras represents the constraints (say $so(p,q)$) and the other coincides with the algebra of observables (as is the case in examples \ref{s5.2} and \ref{s5.3}). The constraints would then impose that the physical Hilbert space has to carry the trivial representation of $so(p,q)$. Now if the trivial representation is included in the decomposition (\ref{hdecomp}) we can adopt as a physical Hilbert space the isotypical component of the trivial representation (i.e. the direct sum of all trivial representations which appear in the decomposition of $\Fl^2(\Rl^{p+q})$ with respect to the group $O(p,q)$)). The above cited theorem ensures that this space carries a  unitary irreducible\footnote{ In general the decomposition with respect to $SO(p,q)$ differs from the decomposition of $O(p,q)$, i.e. if one considers in addition to rotations reflections. One has to take the transformation behavior of vectors in $\Fl^2(\Rl^{p+q})$ under reflections in $O(p,q)$ into account to get uniqueness and irreducibility.} representation of the observable algebra. The scalar product on this Hilbert space is determined by this representation. 

The same holds if we have the $sl(2,\Rl)$ algebra as constraints and $so(p,q)$ as the algebra of observables.

To determine the representation of the observable algebra on the physical Hilbert space the following relation between the (quadratic) Casimirs of the two algebras involved is administrable (see \cite{howe}):
\ba \label{id}
4(-(h^{(p,q)})^2+(n_1^{(p,q)})^2+(n_2^{(p,q)})^2)=-\!\!\!\sum_{j<k\leq p}\!\!\! A_{jk}^2 -\!\!\!\!\!\!\!\sum_{p<j<k\leq p+q} \!\!\!\!\!\! B_{jk}^2 +\!\!\!\!\sum_{j \leq p,\, k>p} \!\!\! \!\! C_{jk}^2 +1-(\tfrac{p+q}{2}-1)^2
\ea
One can check this relation by direct computation.

 Now, what was said above works perfectly well in the case of compact 
gauge groups as for the case of $SO(3)$ in 
\cite{II} but in the examples \ref{s5.2} and \ref{s5.3} the trivial 
representation of the corresponding constraint algebra does not appear in the decomposition (\ref{hdecomp}). (In fact it never appears, if the constraint algebra is $sl(2,\Rl)$.)  

To elaborate on this, we will sketch (following \cite{howeart}) how one can achieve the decomposition (\ref{hdecomp}) using $sl(2,\Rl)$-representation theory. This decomposition is used in examples \ref{s5.2} and \ref{s5.3}.

To begin with, we consider the $p$-fold tensor product $\otimes_{\!p}\omega\simeq\otimes_{\!p}(D(1/2) \oplus D(3/2))$. One can reduce this tensor product by using repeatedly 
\be \label{decom1}
D(l_1)\otimes D(l_2)\simeq\sum_{j =0}^{\infty} D(l_1+l_2+2j)
\ee
from \cite{howe}. Or one uses the above theorem with $q=0$ and reduces rather the regular representation of $O(p)$ on $\Fl^2(\Rl^p)$. This reduction is known to be given by (generalized) spherical harmonics. Via the identity (\ref{id}) one can determine the Casimir of the corresponding $sl(2,\Rl)$ representation. This determines uniquely the representation for $p \geq 2$, since we know from (\ref{decom1}) that there can only appear representations $D(j)$ with $j\geq 1$. (For the case $p=1$ we already have the decomposition $\omega=D(1/2)\oplus D(3/2)$, where $D(1/2)$ and $D(3/2)$ are not being distinguished by the Casimir. But the vectors in these two representations are being distinguished by their transformation behavior under $O(1)$, where $O(1)$ consists just of the reflection $x \mapsto -x$ and the identity.)  
In this way on gets the explicit form of (\ref{hdecomp}) for the case $q=0$ (see \cite{howeart}):
\be \label{decom2}
\omega^{(p,0)}\simeq \sum_{j=0}^\infty \Fh_{p,j} \otimes D(j+p/2) 
\ee
where $\Fh_{p,j}$ is the representation of $O(p)$ defined by the spherical harmonics (for $S^{(p-1)}$) of degree $j$. The dimension  of $\Fh_{p,j}$, in the following denoted by $C_{p,j}$, is finite and is equal to the multiplicity of $D(j+p/2)$ in $\omega^{(p,0)}$. So, if one is just interested in the $sl(2,\Rl)$ structure, one would have 
\be
\omega^{(p,0)}\simeq \sum_{j=0}^\infty C_{p,j} \, D(j+p/2) \q .    
\ee 
The discussion for the tensor product $\omega^{(0,q)}$ is analogous, all representations $D(l)$ are just replaced by $D^*(l)$:
\be 
\omega^{(0,q)}\simeq \sum_{j=0}^\infty C_{q,j} \, D^*(j+q/2) \q .    
\ee 

Therefore, for the complete reduction of $\omega^{(p,q)}$ we have to tackle
\be
\omega^{(p,q)}\simeq \sum_{j,j'=0}^\infty (C_{p,j}\,C_{q,j'})\,D(j+p/2) \otimes D^*(j+q/2)\q,
\ee
i.e. tensor products of the form $D(l_1)\otimes D^*(l_2)$ (for $l_1,l_2$ positive half integers). We will take these from \cite{howeart}: Suppose $l_2\geq l_1$. Then $D(l_1)\otimes D^*(l_2)$ decomposes as
\ba
D(l_1)\otimes D^*(l_2)\simeq \int_{\tfrac{1}{2}}^{\tfrac{1}{2}+i \infty} P(t,\eps) d\mu(t) \oplus \sum_{\stackrel{0\leq 2l<(l_1-l_2-1)}{l \in \Nl}} D(l_1-l_2-2l)
\ea
where  
\begin{alignat}{3}
\eps &=\tfrac{1}{4}  \q &&\text{for} \q && (l_1+l_2) \in \{\tfrac{1}{2}+2n,n\in \Nl\} \nn \\
\eps &= -\tfrac{1}{4}  \q &&\text{for} \q && (l_1+l_2) \in \{\tfrac{3}{2}+2n,n\in \Nl\} \nn \\
\eps &= 0   \q &&\text{for} \q &&(l_1+l_2) \in \{0+2n,n\in \Nl\} \nn \\
\eps &= \tfrac{1}{2}    \q &&\text{for} \q && (l_1+l_2) \in \{1+2n,n\in \Nl\}\q ,
\end{alignat}
and the meausre $d\mu(t)$ is the Plancherel measure on the unitary dual of the double cover of $SL(2,\Rl)$. The reduction for $l_1 \leq l_2$ is obtained by using $D(l_1)\otimes D^*(l_2)\simeq (D^*(l_1)\otimes D(l_2))^*$.

In these decompositions the trivial representation never appears, therefore the Master Constraint Operator $\MCO=h^2 +n_1^2+n_2^2$ never includes zero in its spectrum. 

On the other hand, in the Refined Algebraic Quantization approach one can find trivial representations (for $p,q \geq 2$ and $p+q$ even), see \cite{louko2}. But these trivial representations do not appear in the decomposition of $\omega^{(p,q)}$ as a (continuous) sum of Hilbert spaces. One can find trivial representations if one looks at the algebraic dual $\Phi^*$ of the dense subspace $\Phi$ in $\Fl^2(\Rl^{p+q})$, where $\Phi$ is the linear span of all Fock states.

\subsection{Explicit Calculations for Example \ref{s5.3}}

\label{appneun}

% connection Neunhoeffer and osci basis
%\ba
%2i x =H_- \\
% 2i y_1=D \\
% 2i y_2=H_+\\
%\Omega=x^2-y_1^2-y_2^2=\frac{1}{4}\Fc \,(old?)\\ 
%|j|=|2l|-1,\quad |l|>0 \quad l \in \Zl \\
%eigenval(x)=-ip \\
%eigenvalue(H_-)=2p \quad k=2p
%\ea

Here we will elaborate on example \ref{s5.3} and construct the explicit solutions to the Master Constraint Operator using the results of \cite{neun}. This may provide some hints how to tackle examples, which do not carry such an amount of group structure, as the present one. 

We will start with equation \ref{firstred}, where we achieved the reduction of $\omega^{(2,0)}$ and $\omega^{(0,2)}$. We managed to write the constraints as      
\ba \label{rep1}
H_-&=&A_+^\dagger A_+ +A_-^\dagger A_- -B_+^\dagger B_+ -B_-^\dagger B_- \nonumber\\
H_+&=&-(A_+A_- + A_+^\dagger A_-^\dagger +B_+B_-+B_+^\dagger B_-^\dagger) \nonumber \\
D&=&i(A_+^\dagger A_-^\dagger -A_+ A_- +B_+B_--B_+^\dagger B_-^\dagger) \q.
\ea
The generators of the maximal compact subgroup $O(2)\times O(2)$ of $O(2,2)$ can be written as
\ba
O_{12}&=&u_1p_2-p_1u_2=A_+^\dagger A_+ -A_-^\dagger A_- \nonumber \\
O_{34}&=&\pi_1v_2-v_1\pi_2=-B_+^\dagger B_+ +B_-^\dagger B_- \q.
\ea

 A convenient (ortho-normal) basis in the kinematical Hilbert space 
 $\Fl^2(\Rl^4)$ is given by the Fock states with respect to $A_+,A_-,B_+$ 
and $B_-$ given by
\be
|k_+,k_-,k'_+,k'_->=\frac{1}{\sqrt{k_+ k_-k'_+ k'_-}}(A_+^\dagger)^{k_+}(A_-^\dagger)^{k_-}(B_+^\dagger)^{k'_+}(B_-^\dagger)^{k'_-}|0,0,0,0>
\ee
where $|0,0,0,0>$ is the state which is annihilated by all four annihilation operators and $k_+,k_-,k'_+,k'_-\in \Nl$. These states are eigenstates of $H_-\,$, $O_{12}$ and $O_{34}$ with eigenvalues
\ba
k&:=&\text{eigenval}(H_-)=k_++k_--k'_+-k'_- \nn \\
j&:=&\text{eigenval}(O_{12})=k_+-k_-     \nn \\
j'&:=&\text{eigenval}(O_{34})=-k'_++k'_-    \q. 
\ea
The common eigenspaces $V(j,j')$ of the operators $O_{12}$ and $O_{34}$ are left invariant by the $sl(2,\Rl)$-algebra (\ref{rep1}), since there only appear combinations of $A_+A_-\,$, $B_+B_-$, their adjoints and number operators, which leave the difference between particels in the plus polarization and particels in the minus polarization invariant. Moreover the kinematical Hilbert space is a direct sum of all the (Hilbert) subspaces $V(j,j')$ (since these $V(j,j')$ constitute the spectral decomposition of the self adjoint operators $O_{12}$ and $O_{34}$):
\be
\Fl^2(\Rl^4)=\sum_{j,j'\in \Zl} V(j,j') \q.
\ee
The scalar product on $V(j,j')$ is simply gained by restriction of the $\Fl^2$-scalar product to $V(j,j')$.

The space $V(j,j')$ still carries an $sl(2,\Rl)$- representation, which can be written as a tensor product, where the two factor representations  are     
\begin{xalignat}{2} \label{facs}
H_-^{(A)}&=A_+^\dagger A_+ +A_-^\dagger A_- +1 \q \q &  H_-^{(B)}&=-(B_+^\dagger B_+ +B_-^\dagger B_-+1) \nonumber\\
H_+^{(A)} &=-(A_+A_- + A_+^\dagger A_-^\dagger) \q\q & H_+^{(B)}&=  -(B_+B_-+B_+^\dagger B_-^\dagger)    \nonumber \\
D^{(A)} &=i(A_+^\dagger A_-^\dagger -A_+ A_-)\q\q & D^{(B)}&= i(B_+B_--B_+^\dagger B_-^\dagger) \q.
\end{xalignat}
Each $V(j,j')$ has a basis $\{|k_+,k_-,k'_+,k'_->,k_+-k_-=j \,\wedge\, -k'_++k'_-=j'\}$, which is also an eigenbasis for $H_-^{(A)}$ and $H_-^{(B)}$. Therefore it is easy to check that $H_-^{(A)}$ has on $V(j,j')$ a lowest eigenvalue given by $(|j|+1)$, more generally the spectrum of $H_-^{(A)}$ is non-degenerate and given by $\text{spec}(H_-^{(A)})=\{|j|+1+2n,n \in \Nl\}$. Similarly, $H_-^{(B)}$ has a highest eigenvalue $-(|j|+1)$ on $V(j,j')$ and the spectrum is  $\text{spec}(H_-^{(B)})=\{-(|j|+1+2n),n \in \Nl\}$. From this one can deduce that the representation given on $V(j,j')$ is isomorphic to $D(|j|+1)\otimes D(|j'|+1)$, i.e. a tensor product of a positive discrete series representation and a representation fom the negative discrete series. 

Now, what we want achieve is a spectral composition of the Master Constraint Operator $\MCO$ on each of the subspaces $V(j,j')$. (Clearly, the Master Constraint Operator leaves these spaces invariant.) The Master Constraint Operator is the sum of a multiple of the $sl(2,\Rl)$-Casimir and $2H_-^2$. The latter two operators commute, so we can diagonalize them simultanously. This problem was solved in \cite{neun}. There, another realization of the representation $D(|j|+1)\otimes D(|j'|+1)$ was used, hence to use the results of  \cite{neun}, we have to construct a (unitary) map, which intertwines between our realization and the realization in \cite{neun}. 

To this end, we will depict the realization used in \cite{neun}, at first for representations from the positive and negative discrete series. The Hilbert spaces for these realizations are function spaces on the open unit disc in $\Cl$. For the positive discrete series $D(l),l \in \Nl-{0}$ the Hilbert space, which we will denote by $\Fh_l$, consists of holomorphic functions and for the negativ discrete series $D^*(l)\,,l \in \Nl_{0}$ the Hilbert space ($\Fh_{*l}$) is composed of anti-holomorphic functions. The scalar product is in both cases 
\be
<f,h>_l =\frac{l-1}{\pi}\int_{D} f(z) \overline{h(z)} (1-|z|^2)^{l-2} dx\,dy
\ee  
where $D$ is the unit disc and $dx\,dy$ is the Lebesgue measure on $\Cl$. (For $l=1$ one has to take the limit $l \rightarrow 1$ of the above expression.) An ortho-normal basis is given by
\be
f^{(l)}_n:=(\mu_{l}(n))^{-\tfrac{1}{2}}z^{n}\q  (n \in \Nl)\q  \text{with}\,\, \mu_{l}(n)=\frac{\Gamma(n+1)\Gamma(l)}{\Gamma(l+n)} 
\ee
for the positiv discrete series; for the negative series an ortho-normal basis is
\be
f^{(*l)}_n:=(\mu_{l}(n))^{-\tfrac{1}{2}}\overline{z}^{n}\q  (n \in \Nl)\q . 
\ee
In this realizations the $sl(2,\Rl)$-algebra acts as follows: for the positive discrete series $D(l)$
\ba \label{pos}
H^{(l)}_-&=&l+2z\frac{d}{dz} \nn \\
H^{(l)}_+&=&-lz -(z+z^{-1})z\frac{d}{dz} \nn \\
D^{(l)}&=&ilz+i(z-z^{-1})z\frac{d}{dz}
\ea
and for the negative discrete series $D^*(l)$
\ba \label{neg}
H^{(*l)}_-&=&-l-2\overline{z}\frac{d}{d\overline{z}} \nn \\
H^{(*l)}_+&=&-l\overline{z} +(\overline{z}+\overline{z}^{-1})\overline{z}\frac{d}{d\overline{z}} \nn \\
D^{(*l)}&=&-il\overline{z}+i(\overline{z}-\overline{z}^{-1})\overline{z}\frac{d}{d\overline{z}}\q.
\ea
The aforementioned bases $\{f^{(l)}_n\}$ and $\{f^{(*l)}_n\}$ are eigen-bases for $H^{(l)}_-$ resp. $H^{(*l)}_-$ with eigenvalues $\{l+2n\}$ resp. $\{-l-2n\}$ (where always $n \in \Nl$). 

The representation space of the tensor product $D(l)\otimes D^*(l')$ is the tensor product $\Fh_{l}\otimes \Fh_{*l'}$, which has as an ortho-normal basis $\{ f^{(l)}_n\otimes f^{(*l')}_{n'},\,n,n'\in \Nl\}$. The tensor product representation is obtained by adding the corresponding $sl(2,\Rl)$-representatives from  (\ref{pos}) and (\ref{neg}).  
 
Now, considering the properties of the bases 
$\{|k_+,k_-,k'_+,k'_->,k_+-k_-=j \,\wedge\, -k'_++k'_-=j'\}$ 
for 
$V(j,j')$ and $\{ f^{(|j|+1)}_n\otimes f^{(*(|j'|+1))}_{n'}\}$ 
for 
$\Fh_{|j|+1}\otimes \Fh_{*(|j'|+1)}$ 
it is very suggestive to construct a unitary map between these two Hilbert spaces  by simply matching the bases:
\ba \label{unita}
U:\q \Fh_{|j|+1}\otimes \Fh_{*(|j'|+1)} &\rightarrow & V(j,j') \nn \\
   f^{(|j|+1)}_n\otimes f^{(*(|j'|+1))}_{n'} & \mapsto & (-1)^{n'}|k_+,k_-,k'_+,k'_-> \q \text{where} \nn \\
&& 2 n=k_++k_--|j|\, , \quad j=k_+-k_- \, , \nn \\
&& 2 n'=k'_++k'_--|j'|\, , \quad j=-k'_++k'_-  \q .
\ea
One can check that this map intertwines the $sl(2,\Rl)$-representations. (For this to be the case the factor $(-1)^{n'}$ in (\ref{unita}) is needed.) Since this map maps an orthonormal basis to an orthonormal basis it is an (invertible) isometry and can be continued to the whole Hilbert space (which justifies the notation in (\ref{unita})). We will later use this map to adopt the results of \cite{neun} to our situation.

In the following we will sketch how the spectral decomposition of the Casimir operator in the $D(l)\otimes D^*(l')$-representation is achieved in \cite{neun}. The Casimir operator is
\ba \label{casi}
\Fc &=&\!\!\tfrac{1}{4}(-(H^{(l)}_-+H^{(*l')}_-)^2 + (H^{(l)}_++H^{(*l')}_+)^2
+(D^{(l)}+D^{(*l')})^2) \nn \\
&=&\!\! -(1-z_1\overline{z}_2)^2\partial_{z_1}\partial_{\overline{z}_2}+
l'(1-z_1 \overline{z}_2)z_1 \partial_{z_1} +  
 l(1-z_1 \overline{z}_2)\overline{z}_2\partial_{\overline{z}_2}-
\tfrac{1}{4}(l-l')^2+\tfrac{1}{2}(l+l')-l\, l'z_1\overline{z}_2 \, .\nn \\
\ea
 This operator commutes with all $sl(2,\Rl)$-generators and in particular with $H^{ll'}_-=(H^{(l)}_-+H^{(*l')}_-)$, i.e. it leaves the eigenspaces of $H^{ll'}_-$ invariant. To take advantage of this fact one introduces new coordinates $z=z_1 \overline{z}_2\, ,w=z_1$ and rewrites functions in  $\Fh_{l}\otimes \Fh_{*l'}$ as a Laurent series in $w$ (where the coefficients are functions of $z$). Since functions of the form $f(z)w^{\tfrac{1}{2}(k-l+l')}$ are eigenfunctions of $H^{ll'}_-$ with eigenvalue $k$ one has effectively achieved the spectral decomposition of $H^{ll'}_-$. (The number $\tfrac{1}{2}(k-l+l')$ is always a whole number, since $k$ is (un)even iff $(l-l')$ is (un)even.)  The linear span of all these functions (with fixed $k$) completed with respect to the subspace-topology coming from $\Fh_{l}\otimes \Fh_{*l')}$ is 
% an eigenspace for the self-adjoint operator $H^{ll'}_-$ and therefore
 a Hilbert space, abbreviated by $\Fh(k,l,l')$. Since the power of $w$ is fixed, this Hilbert space is a space of functions of $z$. The scalar product in this Hilbert space is characterized by the fact that
\be
\{(\mu_{l}(n+\tfrac{1}{2}(k-l+l'))\mu_{l'}(n)z^n \q |\q \text{max}(0,\tfrac{1}{2}(-k+l-l')\leq n<\infty \}
\ee
is an orthonormal basis.

One can restrict the Casimir (\ref{casi}) to this Hilbert space (since it leaves the $H^{ll'}_-$-eigenspaces invariant) obtaining
\ba
\Fc_k=(1-z)\bigg( -z(1-z)\frac{d^2}{dz^2}-\tfrac{1}{2}(k-l+l'+2-(k+l+3l'+2)z)\frac{d}{dz} + \nn \\ 
\tfrac{1}{2}l'\,(k+l+l')+\tfrac{1}{4}(l+l')(2-l-l')(1-z)^{-1}\bigg) \q \q.
\ea
Likewise the Master Constraint Operator restricts to $\Fh(k,l,l')$ and can be written as\ba \label{appendixdiffop}
M_k=4 \Fc_k + 2k^2 \q.
\ea
These operators are ordinary second order differenential operators and their spectral decomposition is effected in \cite{neun} by using (modifications of) the Rellich-Titchmarsh-Kodaira-theory. We will not explain this procedure but merely cite the results.

The eigenvalue equation for the master constraint $(M_k-\lambda)f=0$ on $\Fh(k,l=|j|+1,l'=|j'|+1)$ has two linearly independent solutions (since it is a second order differential operator), a near $z=0$ regular solution being

%The master constraint leaves the spaces $V(k,j,j')$ invariant. On these spaces the master constraint equation reduces to an (ordinary) difference equation. In the $sl(2,\Rl)$-realization used in \cite{neun} it is an ordinary differential equation in the variable $z=z_1 \overline{z_2}$.

%The (regular) solution to $(M-\lambda)f=0$ in the latter realization is for 
\ba \label{appendixsol1}
f_{k,j,j'}(z,t)&=&(1-z)^{1-t-\frac{1}{2}(|j|+|j'|+2)}F(1-t+\tfrac{1}{2}(-|j|+|j'|),1-t+\tfrac{1}{2}k,1+\tfrac{1}{2}(k-|j|+|j'|);z)
\ea
for $k-|j|+|j'|\geq 0$
and 
\ba
f_{k,j,j'}(z,t)&=&(1-z)^{1-t-\frac{1}{2}(|j|+|j'|+2)} z^{\frac{1}{2}(-k+|j|-|j'|)} \times \nn \\ &&\times   F(1-t-\tfrac{1}{2}k,1-t+\tfrac{1}{2}(|j|-|j'|),1+\tfrac{1}{2}(-k+|j|-|j'|);z) \q\q\q\q\q\q\q
\ea
for $k-|j|+|j'|\leq 0$, where $t=\tfrac{1}{2}(1+\sqrt{1-\lambda+2k^2}),\, \text{Re}(t) \geq \tfrac{1}{2}$ and $F(a,b,c;z)$ is the hypergeometric function. For $\lambda(k,|j|,|j'|)$ in the spectrum of the Master Constraint Operator these solutions are generalized eigenvectors of the Master Constraint Operator.

The spectrum has a continuous part and a discrete part. 
There is a discrete part only if $k>0$ for $|j|-|j'| \geq 2$ or $k<0$ for $|j|-|j'|\leq 2$.:
\ba
&\lambda_{\text{discr}}=4 t(1-t)+ 2k^2 \geq 2k^2-k^2+2|k| \quad \nn \\
&\text{with} \quad t=1,2,\ldots,\tfrac{1}{2}\text{min}(|k|,||j|-|j'||)\,\, \text{for even} \,\,k  \nonumber \\
&\text{with} \quad t=\tfrac{3}{2},\tfrac{5}{2} \ldots,\tfrac{1}{2}\text{min}(|k|,||j|-|j'||)\,\, \text{for odd} \,\,k  \nonumber \\
&\text{and}\\
&\lambda_{\text{cont}}=1+x^2+2k^2>0 \quad x \in \big[0,\infty\big) \q .
\ea

The spectral resolution of a function $f(z)$ in $\Fh(k,l=|j|+1,l'=|j|+1)$ is for $k-|j|+|j'|\geq 0$
\ba \label{specres1}
f(z)&=&\sum_{\lambda_{\text{discr}}} A(\lambda_{\text{discr}}) +
\int_{\frac{1}{2}}^{\frac{1}{2}+i\infty} (2t-1)\mu(j,j',k,t) <f,f_{k,j,j'}(\cdot,t)>f_{k,j,j'}(z,t)\,dt \nn \\
\mu(j,j',k,t) &=&\frac{1}{i \pi^2\G(|j|+1)\G(|j'|+1)\G^2(\tfrac{1}{2}(k-|j|+|j'|+2)) } \sin \pi t\, \cos \pi t\,\,\times \nn \\
 && \times \, |\G(t+\tfrac{1}{2}k)\G(t-\tfrac{1}{2}(2-|j|-|j'|))|^2|\G(t-\tfrac{1}{2}(|j|-|j'|))|^2
\ea
and for $k-|j|+|j'|\leq 0$
\ba\label{specres2}
f(z)&=&\sum_{\lambda_{\text{discr}}} B(\lambda_{\text{discr}}) +               \int_{\frac{1}{2}}^{\frac{1}{2}+i\infty} (2t-1)\mu(j,j',k,t) <f,f_{k,j,j'}(\cdot,t)>f_{k,j,j'}(z,t)\,dt  \nn \\
\mu(j,j',k,t) &=& \frac{1}{i \pi^2\G(|j|+1)\G(|j'|+1)\G^2(\tfrac{1}{2}(-k+|j|-|j'|+2)) }  \sin \pi t\, \cos \pi t\,\,\times \nn \\
&& \times \, |\G(t-\tfrac{1}{2}k)\G(t-\tfrac{1}{2}(2-|j|-|j'|))|^2|\G(t+\tfrac{1}{2}(|j|-|j'|))|^2
\ea
where in the following we do not need $A(\lambda_{\text{discr}})$ and $B(\lambda_{\text{discr}})$ in explicit form. %(The scalar product $<\cdot,\cdot>$  on $\Fh(k,|j|+1,|j'|+1)$ is the induced one from $\Fh_{|j|+1}\otimes \Fh_{*(|j'|+1))}$.) 
This gives the following resolution of a function $f(z_1,\overline{z}_2)$ in $\Fh_{|j|+1}\otimes \Fh_{*(|j'|+1))}$:
\ba
f(z_1,\overline{z}_2)=\text{discr.\;part}+ \!\!\sum_{k}\int_{\frac{1}{2}}^{\frac{1}{2}+i\infty}\!\!\!\!\!\!\!\!\! (2t-1)\mu(j,j',k,t) 
<f,f_{k,j,j'}(\cdot,\cdot,t)>f_{k,j,j'}(z_1,\overline{z}_2,t)\,dt  
\ea
where
\be  \label{solu1}
f_{k,j,j'}(z_1,\overline{z}_2,t)=f_{k,j,j'}(z_1\overline{z}_2,t)\,z_1^{\tfrac{1}{2}(k-|j|+|j'|)},
\ee
and the sum is over all whole numbers $k$ with the same parity as $(j-j')$.

Now we can use the map $U$ in (\ref{unita}) to transfer these results to the subspaces $V(j,j')$ of the kinematical Hilbert space $\Fl^2(\Rl^4)$. To this end we rewrite (\ref{solu1}) into a power series in $z_1$ and $\overline{z}_2$ using the definition of the hypergeometric function
\be
F(a,b,c\,;z)=\frac{\G(c)}{\G(a)\G(b)}\sum_{n=0}\frac{\G(a+n)\G(b+n)}{\G(c+n)\G(1+n)}z^n
\ee
and
\be
(1-z)^{1-d}=\sum_{n=0}\frac{\G(d+k-1)}{\G(d-1)\G(k+1)}z^k \q.
\ee

For
$k-|j|+|j'|\geq 0$ we obtain
\be  \label{sol1a}
f(t;k,j,j')=U(f_{k,j,j'}(z_1,\overline{z}_2))=\sum_{m=0} a_m \,|k_+(m),k_-(m),k'_+(m),k'_-(m)>
\ee
where
\begin{xalignat}{2}  \label{kkkk}
k_+ &=m+\tfrac{1}{2}(k+j+|j'|) & k_-&=m+\tfrac{1}{2}(k-j+|j'|) \nn \\
k'_+&=m+\tfrac{1}{2}(|j'|-j') & k'_-&=m+\tfrac{1}{2}(|j'|+j')
\end{xalignat}
%(Since we have the relation $(-1)^k=(-1)^j(-1)^{j'}$, the numbers $k_+,k_-,k'_+,k'_-$ are natural numbers.)
and 
\ba \label{am}
a_m &=&(-1)^m (\mu_{(|j|+1)}(m+\frac{1}{2}(k-|j|+|j'|) ))^{\tfrac{1}{2}} (\mu_{(|j'|+1)}(m ))^{\tfrac{1}{2}}    \,\, \times \nn \\
&& \times \frac{\G(1+\tfrac{1}{2}(k-|j|+|j'|) )}{\G(1-t+\tfrac{1}{2}(-|j|+|j'|) )\G(1-t+\tfrac{1}{2}k )} \,\, \times\nonumber \\
&& \times\,\sum_{l=0}^m \frac{\G(1-t+\tfrac{1}{2}(-|j|+|j'|)+l ) \G(1-t+\tfrac{1}{2}k+l )}{\G(1+\tfrac{1}{2}(k-|j|+|j'|)+l )\G(1+l )}\frac{\G(t+\frac{1}{2}(|j|+|j'|) +(m-l))}{\G(m-l+1)\G(t+\frac{1}{2}(|j|+|j'|) )}\q .\nn \\
\ea
For $k-|j|+|j'|\leq 0$ the coefficient $a_m$ in (\ref{sol1a}) is obtained from $(\ref{am})$ by replacing $k$ with $-k$, switching $|j|$ and $|j'|$ and multiplying with $(-1)^{\tfrac{1}{2}(-k+|j|-|j'|)}$.

We could use the vectors $f(t;k,j,j')$ to construct the spectral decomposition of $\Fl^2(\Rl^4)$. However, we want to achieve a spectral measure, which is independent of $k,j$ and $j'$. For this purpose we normalize the solutions (\ref{sol1a}) to
\be
|t,k,j,j'> =  \bigg(i\,\frac{ \mu(j,j',k,t)}{\sin \pi t \, \cos \pi t } \bigg)^{\tfrac{1}{2}}f(t;k,j,j')\q.
\ee
Now we can decompose a vector $|f>\in \Fl^2(\Rl^4)$ as follows
\ba
|f>&=&\text{discrete\; part}+\nn \\
 &&\sum_{k,j,j'}\int_{\tfrac{1}{2}}^{\tfrac{1}{2}+i\infty} i\,(1-2t) \sin \pi t \, \cos \pi t\, <f|t,k,j,j'>|t,k,j,j'> dt
\ea
where the sum is over all whole numbers $k,j,j'$ with $(-1)^k=(-1)^{j-j'}$.

From this it follows, that $\Fl^2(\Rl^4)$ decomposes into a direct sum (for the discrete part) and direct integral of Hilbert spaces $\Fh(t)$, where in each $\Fh(t)$ an ortho-normal basis is given by the vectors $|t,k,j,j'>$. As explained in section \ref{s5.3} our physical Hilbert space $\Fh''$ consists of vectors with $t=\tfrac{1}{2},k=0$ and $|j|=|j'|$. In this case these vectors are given by
\ba \label{finstates}
|j,j'>&=&|t=\tfrac{1}{2},k=0,j,j'>\q =\sum b_m \,|k_+(m),k_-(m),k'_+(m),k'_-(m)> \nn \\
b_m &=&(-1)^m   \frac{\Gamma(m+1)}{\Gamma(|j|+1+m)}
 \sum_{l=0}^m \frac{(\G(\tfrac{1}{2}+l ))^2}{(\G(1+l ))^2}\frac{\G(|j|+\tfrac{1}{2} +(m-l))}{\G(m-l+1)}
\ea
with $k_+(m),k_-(m),k'_+(m),k'_-(m)$ given by (\ref{kkkk}).

Now we want to check our results by calculating the action of the Master Constraint Operator and of the observables on the states (\ref{finstates}).

The Master Constraint Operator rewritten in terms of annihilation and creation operators is
\ba
\MCO&=&2\big(2N(A_+)N(A_-)+N(A_+)+N(A_-)+2N(B_+)N(B_-)+\nonumber \\
 & & \q N(B_+)+N(B_-)+2+ 
2A_+^\dagger A_-^\dagger B_+^\dagger B_-^\dagger+2A_+A_-B_+B_- \big)+\nonumber \\ &&\big(N(A_+)+N(A_-)-N(B_+)-N(B_-)\big)^2
\ea
where $N(i)$ stands for the number operator for quanta of type $i$.

%We make the ansatz
%\be
%|\lambda,k,j,j'>=\sum_{m=0} a_m \,|k_+(m),k_-(m)>\otimes|k'_+(m),k'_-(m)>
%\ee
%with
%\ba
%k_+&=&m+\tfrac{1}{2}(k+j+|j'|) \\
%k_-&=&m+\tfrac{1}{2}(k-j+|j'|) \\
%k'_+&=&m+\tfrac{1}{2}(|j'|-j') \\
%k'_-&=&m+\tfrac{1}{2}(|j'|+j')
%\ea
%for the eigensolutions of the master constraint and obtain the following equation
%for the coefficients $a_m$: 
%\ba
%0=&(4(k_+(m)k_-(m)+k'_+(m)k'_-(m))+2(k_+(m)+k_-(m)+k'_+(m)+k'_-(m)+2)+k^2-\lambda)\,a_m +\nonumber \\
%&\sqrt{k_+(m)k_-(m)k'_+(m)k'_-(m)}\,a_{m-1}+\nonumber \\ 
%&\sqrt{(k_+(m)+1)(k_-(m)+1)(k'_+(m)+1)(k'_-(m)+1)}\,a_{m+1}
%\ea

The eigenvalue equation $(M-\lambda)|j,j'>=0$ for the states (\ref{finstates}) can be written as an equation for the coefficients $b_m$: 
\ba \label{appendixmasterequ}
0=(8(m+|j|+1)m +4|j|+4-\lambda)\,b_m+4m(m+|j|)\,b_{m-1}+
4(m+1)(m+|j|+1)\,b_{m+1} \nn \\
\ea 
(The coeffecient $b_{-1}$ is defined to be zero.)
One can check, that the coefficients (\ref{finstates}) fulfill this equation for $\lambda=1$: For this purpose one introduces
\be \label{appendtildebm}
\tilde b_m=(-1)^m \frac{\Gamma(|j|+1+m)}{\Gamma(m+1)}b_m= \sum_{l=0}^m \frac{(\G(\tfrac{1}{2}+l ))^2}{(\G(1+l ))^2}\frac{\G(|j|+\tfrac{1}{2} +(m-l))}{\G(m-l+1)}
\ee
and realizes that the $\tilde b_m$'s are the coefficients in the power expansion of the function $\Gamma(|j|+1/2)\,f_{0,j,j'}(z,t=\tfrac{1}{2})$  (with $|j|=|j'|$) from (\ref{appendixsol1}). This function fullfills the differential equation $M_0 \cdot f =f$ where $M_{k=0}$ is the differential operator from (\ref{appendixdiffop}). One can rewrite the differential equation for $f$ into a equation for the coefficients $\tilde b_m$ in a power expansion for $f$. If one replaces $\tilde b_m$ with $b_m$ according to the first part of equation (\ref{appendtildebm}) one will get equation (\ref{appendixmasterequ}). Therefore the coefficients $b_m$ fulfill this equation.

The observables can be written as
\ba
Q_1 &=&\tfrac{i}{2}(A_+B_+-A_-B_--A_+^\dagger B_+^\dagger+A_-^\dagger B_-^\dagger) \nonumber \\
Q_2 &=& \tfrac{-1}{2}(A_+B_++A_-B_-+A_+^\dagger B_+^\dagger+A_-^\dagger B_-^\dagger) \nonumber \\
Q_3 &=& \tfrac{1}{2}(N(A_+)-N(A_-)+N(B_+)-N(B_-)) \nonumber \\
P_1 &=& \tfrac{i}{2}(A_+B_--A_+^\dagger B_-^\dagger-A_-B_++A_+^\dagger B_+^\dagger) \nonumber \\
P_2 &=& \tfrac{-1}{2}(A_+B_-+A_+^\dagger B_-^\dagger+A_-B_++A_+^\dagger B_+^\dagger) \nonumber \\
P_3 &=& \tfrac{1}{2}(N(A_+)-N(A_-)-N(B_+)+N(B_-)) \nonumber \\
\\
Q_+ &=& \tfrac{1}{\sqrt{2}}(Q_1+iQ_2)=\tfrac{-i}{\sqrt{2}}(A_+^\dagger B_+^\dagger +A_-B_-) \nonumber \\
Q_- &=& \tfrac{1}{\sqrt{2}}(Q_1-iQ_2)=\tfrac{+i}{\sqrt{2}}(A_-^\dagger B_-^\dagger +A_+B_+) \nonumber \\
P_+ &=& \tfrac{1}{\sqrt{2}}(P_1+iP_2)=\tfrac{-i}{\sqrt{2}}(A_+^\dagger B_-^\dagger +A_-B_+) \nonumber \\
P_- &=& \tfrac{1}{\sqrt{2}}(P_1-iP_2)=\tfrac{+i}{\sqrt{2}}(A_-^\dagger B_+^\dagger +A_+B_-) \q.
\ea
In section \ref{s5.3} we concluded that on the physical Hilbert space $\Fh''$ the observable algebra is generated by operators of the form $\Theta(Q_3)Q_i\Theta(Q_3)$ and $\Theta(P_3)P_i\Theta(P_3)$. Therefore we will just depict the action of $Q_\pm$ on states with zero $P_3$-eigenvalue and of $P_\pm$ on states with zero $Q_3$-eigenvalue. One can determine from this the action of the observable algebra on $\Fh''$.

To begin with, we consider the action of $Q_+$ on states $|j,-j>,\, j\geq 0$, i.e. states with $P_3$-eigenvalue zero and nonnegative $Q_3$-eigenvalue:
\ba
|j,-j>&=&\sum_{m=0} b_m(j)  |m+j,m,m+j,m>  \nn \\
b_m(j)&=& (-1)^m   \frac{\Gamma(m+1)}{\Gamma(|j|+1+m)}
  \sum_{l=0}^m \frac{(\G(\tfrac{1}{2}+l ))^2}{(\G(1+l ))^2}\frac{\G(|j|+\tfrac{1}{2} +(m-l))}{\G(m-l+1)} \q .
\ea
On these states $Q_+$ acts as
\ba
Q_+|j,-j>
    %=&\tfrac{-i}{\sqrt{2}}\bigg(\sum_{m=0} (m+j+1)b_m\,|m+j+1,m>\otimes|m+j+1,     m>+ \nn \\ 
     %&\sum_{m=1}  mb_m\,|m+j,m-1>\otimes|m+j,m-1>}\bigg) \nonumber \\
&=& \tfrac{-i}{\sqrt{2}}\sum_{m=0}((m+1)b_{m+1}(j)+ (m+j+1)b_m(j)) 
\,\, |m+j+1,m,m+j+1,m>  \nn \\
& \underset{(*)}{=}&\tfrac{-i}{\sqrt{2}}(j+\tfrac{1}{2})\, \, |j+1,-(j+1)>\q .
\ea
For $j<0$ we have
\ba
Q_+|j,-j>&=&\tfrac{-i}{\sqrt{2}}\sum_{m=0}(m\,b_{m-1}(j)+ (m+|j|)b_m(j))
\,\, |m,m+|j|-1,m,m+|j|-1> \nn \\
&\underset{(*)}{=}&\tfrac{i}{\sqrt{2}}(j+\tfrac{1}{2})\, \, |j+1,-(j+1)>\q .
\ea
For the equalities marked with a star $(*)$ we have to check the relations 
\ba \label{appendixrelat}
(m+1)b_{m+1}(j)+ (m+j+1)b_m(j) &=& (j+\tfrac{1}{2})b_m (j+1) \q \text{for} \q j\geq 0\nn \\
m\,b_{m-1}(|j|)+ (m+|j|)b_m(j) &=& (|j|-\tfrac{1}{2})b_m (|j|-1)\q \text{for} \q j< 0 \q .
\ea
The last equation is verified by using the $\tilde b_m(j)$'s defined in (\ref{appendtildebm}), which are the coefficients of 
\be
f_{|j|}(z):=\Gamma(|j|+1/2)\,f_{0,j,\pm j}(z,t=\tfrac{1}{2})=\Gamma(|j|+1/2) (1-z)^{-|j|-1/2}F(1/2,1/2,1;z)
\ee
in a power expansion in $z$. Then, rewriting of the identity $(1-z)f_{|j|}(z)= (|j|-1/2)\,f_{|j|-1}(z)$ into an equation for the $\tilde b_m(j)$ and furthermore for the $b_m(j)$ results in the last equation of  (\ref{appendixrelat}). For the first equation one starts with the differential equation $M_0 \cdot f_{|j|}(z)=f_{|j|}(z)$ and replaces there $(1-z)^{-1}f_{|j|}(z)$ with $(|j|+1/2)^{-1}f_{|j|+1}(z)$. This then translates into the first equation of (\ref{appendixrelat}) for the coefficients $b_m(j)$.

The relations (\ref{appendixrelat}) will also ensure the following equalities:
\ba
Q_-|j,-j>&=&\tfrac{i}{\sqrt{2}}(j-\tfrac{1}{2})|j-1,-(j-1)> \nn \\
P_+|j,j>&=&\tfrac{-i}{\sqrt{2}}(j+\tfrac{1}{2})|j+1,j+1> \nn \\
P_-|j,j>&=&\tfrac{i}{\sqrt{2}}(j-\tfrac{1}{2})|j-1,j-1> \q .
\ea
These formulas differ by  phase factors from the formulas in \ref{physhilbert}. One can adjust these phase factors to one by choosing a new basis $|j,\eps j>'=(-i)^j |j,\eps j>$. Therefore the results of this section and section \ref{s5.3} are consistent.

\end{appendix}


\begin{thebibliography}{99}

\parskip -5pt


\bibitem{7.0} T. Thiemann, ``The Phoenix Project: Master Constraint 
Programme for Loop Quantum Gravity'', gr-qc/0305080

\bibitem{I} B. Dittrich, T. Thiemann, ``Testing the Master Constraint 
Programme for Loop Quantum Gravity I. General Framework'',
gr-qc/0411138

\bibitem{II} B. Dittrich, T. Thiemann, ``Testing the Master Constraint 
Programme for Loop Quantum Gravity II. Finite Dimensional Systems'',
gr-qc/0411139

\bibitem{IV} B. Dittrich, T. Thiemann, ``Testing the Master Constraint 
Programme for Loop Quantum Gravity IV. Free Field Theories'',
gr-qc/0411141

\bibitem{V} B. Dittrich, T. Thiemann, ``Testing the Master Constraint 
Programme for Loop Quantum Gravity V. Interacting Field Theories'',
gr-qc/0411142

\bibitem{1.1} 
C. Rovelli, ``Loop Quantum Gravity", Living Rev. Rel. {\bf 1} (1998) 1,
gr-qc/9710008\\
T. Thiemann,``Lectures on Loop Quantum Gravity'', Lecture Notes in 
Physics, {\bf 631} (2003) 41 -- 135, gr-qc/0210094\\
A. Ashtekar, J. Lewandowski, ``Background Independent Quantum Gravity:
A Status Report'', Class. Quant. Grav. {\bf 21} (2004) R53; 
[gr-qc/0404018]\\
L. Smolin, ``An Invitation to Loop Quantum Gravity'', hep-th/0408048

\bibitem{7.2} C. Rovelli, ``Quantum Gravity'', Cambridge University Press,
Cambridge, 2004

\bibitem{7.3} T. Thiemann, ``Modern Canonical Quantum General 
Relativity'', Cambridge University Press, Cambridge, 2005,
gr-qc/0110034

\bibitem{7.1} T. Thiemann, ``Anomaly-free Formulation of non-perturbative,
four-dimensional Lorentzian Quantum Gravity", Physics Letters {\bf B380}
(1996) 257-264, [gr-qc/9606088]\\
T. Thiemann, ``Quantum Spin Dynamics (QSD)",
Class. Quantum Grav. {\bf 15} (1998) 839-73, [gr-qc/9606089];
``II. The Kernel of the Wheeler-DeWitt Constraint Operator",
Class. Quantum Grav. {\bf 15} (1998) 875-905, [gr-qc/9606090];
``III.
Quantum Constraint Algebra and Physical Scalar Product in Quantum General
Relativity", Class. Quantum Grav. {\bf 15} (1998) 1207-1247,
[gr-qc/9705017];
``IV. 2+1 Euclidean Quantum Gravity as a model to test 3+1
Lorentzian Quantum Gravity", Class. Quantum Grav. {\bf 15} (1998) 
1249-1280, [gr-qc/9705018]; 
``V. Quantum Gravity as the Natural Regulator of the Hamiltonian 
Constraint
of Matter Quantum Field Theories",
Class. Quantum Grav. {\bf 15} (1998) 1281-1314, [gr-qc/9705019];
``VI. Quantum Poincar\'e Algebra and a Quantum Positivity of Energy
Theorem for Canonical Quantum Gravity",
Class. Quantum Grav. {\bf 15} (1998) 1463-1485, [gr-qc/9705020];
``Kinematical Hilbert Spaces for Fermionic and
Higgs Quantum Field Theories",
Class. Quantum Grav. {\bf 15} (1998) 1487-1512, [gr-qc/9705021]


\bibitem{8.2} 
A. Ashtekar, J. Lewandowski, D. Marolf, J. Mour\~ao, T.
Thiemann, ``Quantization for diffeomorphism invariant theories
of connections with local degrees of freedom", Journ. Math. Phys.
{\bf 36} (1995) 6456-6493, [gr-qc/9504018]

\bibitem{7.4} H. Sahlmann, ``When do Measures on the Space of Connections
Support the Triad Operators of Loop Quantum Gravity?'', gr-qc/0207112;
``Some Comments on the Representation Theory of the Algebra Underlying
Loop Quantum Gravity'', gr-qc/0207111\\
H. Sahlmann, T. Thiemann, ``On the Superselection Theory of
the Weyl Algebra for Diffeomorphism Invariant Quantum Gauge Theories'',
gr-qc/0302090;
``Irreducibility of the Ashtekar-Isham-Lewandowski Representation'',
gr-qc/0303074\\
A. Okolow, J. Lewandowski, ``Diffeomorphism Covariant
Representations of the Holonomy Flux Algebra'', gr-qc/0302059

%\bibitem{7.5} 
%A. Ashtekar, C.J. Isham, ``Representations of the Holonomy
%Algebras of Gravity and Non-Abelean Gauge Theories",
%Class. Quantum Grav. {\bf 9} (1992) 1433, [hep-th/9202053]\\
%A. Ashtekar, J. Lewandowski, ``Representation
%theory of analytic Holonomy $C^\star$ algebras", in ``Knots and
%Quantum Gravity", J. Baez (ed.), Oxford University Press, Oxford 1994

%\bibitem{1.6} R. Haag, ``Local Quantum Physics", 2nd ed., Springer 
%Verlag,
%Berlin, 1996

%\bibitem{1.8} 
% M. Bojowald, ``Loop Quantum Cosmology. I. Kinematics",
%Class. Quantum Grav. {\bf 17} (2000) 1489 [gr-qc/9910103];
%``II. Volume Operators",
%Class. Quantum Grav. {\bf 17} (2000) 1509 [gr-qc/9910104];
%``III. Wheeler-DeWitt Operators",
%Class. Quant. Grav. 18:1055-1070,2001 [gr-qc/0008052];
%``IV. Discrete Time Evolution"
%Class. Quant. Grav. 18:1071-1088,2001 [gr-qc/0008053]


\bibitem{1.10} D. Giulini, D. Marolf , ``A Uniqueness Theorem for 
Constraint 
Quantization"
Class. Quant. Grav. 16:2489-2505,1999, [gr-qc/9902045];
``On the Generality of Refined Algebraic
Quantization", Class. Quant. Grav. 16:2479-2488,1999, [gr-qc/9812024]

\bibitem{2.1} J. Klauder, ``Universal Procedure for Enforcing Quantum 
Constraints'', Nucl.Phys.B547:397-412,1999, [hep-th/9901010];
``Quantization of Constrined Systems'', Lect. Notes Phys. 
{\bf 572} (2001) 143-182, [hep-th/0003297]\\
A. Kempf, J. Klauder, ``On the Implementation of Constraints through 
Projection Operators'', J. Phys. {\bf A34} (2001) 1019-1036,
[quant-ph/0009072]


%\bibitem{2.2}  I. M. Gel'fand, N. Ya. Vilenkin, ``Generalized Functions",
%            vol. 4, Applications of Harmonic Analysis, Academic Press,
%            New York, London, 1964

%\bibitem{6.1} J.E. Marsden, P.R. Chernoff, ``Properties of Infinite
%Dimensional Hamiltonian Systems", Lecture Notes in Mathematics,
%Springer-Verlag, Berlin, 1974

\bibitem{8.1} T. Thiemann, ``Quantum Spin Dynamics (QSD): VIII.
The Master Constraint'', in preparation


\bibitem{8.3} H. Sahlmann, T. Thiemann, ``Towards the QFT on
Curved Spacetime Limit of QGR. 1. A General Scheme'', [gr-qc/0207030];
``2. A Concrete Implementation", [gr-qc/0207031]

%\bibitem{8.4} E. Witten, Nucl. Phys. {\bf B311} (1988) 46

\bibitem{8.5} T. Thiemann, ``Quantum Spin Dynamics (QSD): VII.
Symplectic Structures and Continuum Lattice Formulations of
Gauge Field Theories", Class.Quant.Grav.18:3293-3338,2001,
[hep-th/0005232]; ``Gauge Field Theory Coherent States (GCS): I.
General Properties", Class.Quant.Grav.18:2025-2064,2001, [hep-th/0005233];
``Complexifier Coherent States for Canonical Quantum General Relativity", 
gr-qc/0206037\\
T. Thiemann, O. Winkler, ``Gauge Field Theory Coherent States
(GCS): II. Peakedness Properties", Class.Quant.Grav.18:2561-2636,2001,
[hep-th/0005237]; ``III. Ehrenfest Theorems",
Class. Quantum Grav. {\bf 18} (2001) 4629-4681, [hep-th/0005234];
``IV. Infinite Tensor Product and Thermodynamic Limit",
Class. Quantum Grav. {\bf 18} (2001) 4997-5033, [hep-th/0005235]\\
H. Sahlmann, T. Thiemann, O. Winkler, ``Coherent States for
Canonical Quantum General Relativity and the Infinite Tensor Product
Extension", Nucl.Phys.B606:401-440,2001
[gr-qc/0102038]

\bibitem{RS} M. Reed, B. Simon, ``Functional Analysis'', vol. 1, Academic 
Press, New York, 1980

%\bibitem{Thasymp} T. Thiemann,''Generalized Boundary Conditions for 
%General Relativity for the Asymptotically Flat Case in Terms of 
%Ashtekar's Variables'', Class. Quant. Grav. {\bf 12}, 181-198, (1995), 
%gr-qc/9910008

%\bibitem{AshRovLinGr} A. Ashtekar, C. Rovelli, L. Smolin, ``Gravitons and 
%Loops'', Phys. Rev. {\bf 44}, 1740-1755, (1991), hep-th/9202054

%\bibitem{AshLee} A. Ashtekar, J. Lee, ``Weak Field Limit of General 
%Relativity: A New Hamiltonian Formulation'', Int. J. Mod. Phys. {\bf D3}, 
%675-693 (1994) 

%\bibitem{9.1} Rovelli,Montesinos,Thiemann,Louko,Trunk ......
%probably already referred to in earlier sections ************

\bibitem{9.2} A. Paterson, ``Amenability'', American Mathematical Society,
Providence, Rhode Island, 1988\\
J.P. Pier, ``Amenable Locally Compact Groups'', New York, Wiley, 1984\\
A. Carey, H. Grundling, ``Amenability of the Gauge Group'', 
math-ph/0401031


%\bibitem{ReedSimonI}Reed,Simon I
%\bibitem{ReedSimonII}Reed,Simon II
%\bibitem{SchroedOp} G. Teschl: Schr\"odinger Operators

\bibitem{howe} R. Howe, E.C. Tan, ``Non-Abelian Harmonic Analysis -- Applications of SL(2,R)'', (Springer, New-York,1992)

\bibitem{howeart} R. Howe, ``On some results of Strichartz and of Rallis and Schiffman'', J. Funct. Anal. {\bf 32},297 (1979)

\bibitem{howephys} R. Howe, ``Dual pairs in physics: Harmonic oscillators, photons, electrons, and singletons'', Lectures in Appl. Math., Vol. 21, pp.275-286, Amer. Math. Soc., Providence, RI, 1979

\bibitem{neun} H. Neunh\"offer, ``\"Uber Kronecker-Produkte irreduzibler Darstellungen von $SL(2,\Rl)$'', Sitzungsberichte der Heidelberger Akademie der Wissenschaften, Mathematisch-naturwissenschaftliche Klasse, Jahrgang 1978, 167 

\bibitem{rov} M. Montesinos, C. Rovelli, T. Thiemann, ``$SL(2,\Rl)$ model with two Hamiltonian constraints'', Phys. Rev. {\bf D60},044009 (1999), gr-qc/9901073

 \bibitem{louko} J. Louko, C. Rovelli, ``Refined algebraic quantization in 
 the oscillator representation of $SL(2,\Rl)$'', J. Math. Phys. {\bf 41}, 
132 (2000), gr-qc/9907004 

\bibitem{louko2} J. Louko, A. Molgado, ``Group averaging in the $(p,q)$ oscillator representation of $SL(2,\Rl)$'',J. Math. Phys. {\bf 45} 1919 (2004), gr-qc/0312014

\bibitem{trunk} M. Trunk, ''An ``$SL(2,\Rl)$ model of constrained systems: Algebraic constrained quantization'', University of Freiburg preprint THEP 99/3, hep-th/9907056

\bibitem{gamb} R. Gambini, R.A. Porto, ``Relational time in generally covariant quantum systems: Four models'', Phys. Rev. {\bf D63}, 105014 (2001), gr-qc/0101057

\bibitem{repka} J. Repka, ``Tensor products of unitary representations of $SL(2,\Rl)$'', Amer. J. Math. {\bf 100(4)}, 747 (1978)

\bibitem{gomberoff} A. Gomberoff, D. Marolf, ``On Group Averaging for $SO(n,1)$'',Int. J. Mod. Phys. {\bf D8}, 519 (1999), gr-qc/9902069

\bibitem{adams} B.G. Adams, J. Cizek, J. Paldus, ``Lie Algebraic Methods and Their Applications to Simple Quantum Systems'', Adv. Quantum Chem. {\bf 19}, 1 (1988)

\bibitem{bargmann} V. Bargmann, ``Irreducible unitary representations of the Lorentz group'', Ann. of Math. {\bf 48}, 568 (1947)
\end{thebibliography}
\end{document}